\documentclass{svjour2}                    
\smartqed  
\usepackage{graphicx}
\usepackage{amssymb}
  \usepackage{color}
  \usepackage{float}
  \usepackage[numbers,sort&compress]{natbib}

\journalname{myjournal}
\begin{document}

\title{Systemic Losses Due to Counter Party Risk in a Stylized Banking System
}


\author{Annika Birch         \and
        Tomaso Aste 
}


\institute{Annika Birch \at
              Department of Computer Science, University College London, Gower Street, London, WC1E 6BT, UK and Systemic Risk Centre, London School of Economics and Political Sciences, London, WC2A2AE, UK \\
              Tel.: +44-20-76790304\\
              \email{a.birch.11@ucl.ac.uk}           
           \and
           Tomaso Aste \at
           Department of Computer Science, University College London, Gower Street, London, WC1E 6BT, UK and Systemic Risk Centre, London School of Economics and Political Sciences, London, WC2A2AE, UK  \\
              +44-20-76790430\\
              \email{t.aste@ucl.ac.uk}  
}

\date{Received: date / Accepted: date}

\maketitle

\begin{abstract}
We report a study of a stylized banking cascade model investigating systemic risk caused by counter party failure using liabilities and assets to define banks' balance sheet. 
In our stylized system, banks can be in two states: normally operating or distressed and the state of a bank changes from normally operating to distressed whenever its liabilities are larger than the banks' assets. 
The banks are connected through an interbank lending network and, whenever a bank is distressed, its creditor cannot expect the loan from the distressed bank to be repaid, potentially becoming distressed themselves. 
We solve the problem analytically for a homogeneous system and test the robustness and generality of the results with simulations of more complex systems. 
We investigate the parameter space and the  corresponding distribution of operating banks mapping the conditions under which the whole system is stable or unstable. 
This allows us to determine how financial stability of a banking system is influenced by regulatory decisions, such as leverage; we discuss the effect of central bank actions, such as quantitative easing and we determine the cost of rescuing a distressed banking system using re-capitalisation. Finally, we estimate the stability of the UK and US banking systems in the years 2007 and 2012 showing that both banking systems were more unstable in 2007 and connectedness on the interbank market partly caused the banking crisis.
\end{abstract}

\section{Introduction}

During the global financial crisis that started around 2008, it became evident that the structure of the modern financial system can cause sever danger in the event of distress of single banks by spreading the distress through claims on the interbank market to other banks. The risk that banks impose on others through interconnectedness is called counter party risk \cite{Upper2011} and it is the subject of this paper. 

In this study, we are reporting how interconnectedness via the interbank market and the ratios between liabilities and assets  influence the stability of a stylized banking system. The risk of banks to cause system failure is called systemic risk \cite{Fouque2013}. Extending a simple Merton model of default \cite{Merton1973}, we are able to test the systemic resilience of the financial system based on balance sheet quantities and determine ratios at which counter party risk can cause the entire system to fail. 

There is a rich literature on stylized banking models. For instance, in \citep{Battiston2012debtrank, Gai2007, Nier2007, May2010} counter party risk exposed via interbank lending is investigated. Other studies by \cite{Tsatskis2012, Caccioli2012, Solorzano2013}, considered default cascade from an initial shock on asset prices and study market risk of correlated asset classes. These models have been used to investigate avalanches, loss distributions and parameter influences on the stability of the system. Most of the models are simulation based, and use as an initial shock an arbitrary failure of a portion of banks, or arbitrary loss on the value of assets.  

In this paper, we propose a model combing the balance sheet based model, used by \cite{Gai2007} and \cite{Nier2007}, with the contagion model used by \cite{Solorzano2013} creating a stylized banking system that is analogous to the random field Ising model, a well-known model in the statistical physics literature. 
The application of this kind of model in the context of economic and financial behaviour has been reviewed in \cite{Bouchaud2012}, and its application to credit default models has been discussed in \cite{Heise2012}. 

In our stylized banking system, a bank is considered insolvent, if its liabilities are larger than the bank's assets, the so-called balance sheet test  \cite{Goode2010}.
Such insolvency of a bank can be triggered by a random event (such as changes in the value of the assets). 
The interconnectedness between institutions, in form of loans from one institution to another, can propagate this insolvency from a bank to another creating further insolvencies, bringing down -eventually- the entire system. 
In this paper, we discuss a solution of this model, obtained by homogenizing the system. This allows for a mean-field assumption enabling us to compute the equilibrium fraction of surviving banks given changes in the values of the balance sheet quantities.
Further, we test our results changing the structure of the exposure network testing robustness and generality of the mean-field solution. 
We detail the parameter ranges that lead to a stable or unstable system, allowing us to determine restricting ratios between liabilities and assets to ensure a stable banking system. 
Further, we quantify the costs of potential rescue attempts to re-direct an unstable system into a stable region. We find that interbank lending can increase the stability of a banking system but this at the price of an increasing risk of a sudden systemic failure with inflating recovery cost. Finally, we show using balance sheet data for 2007 and 2012 that the US and UK banking system in 2007 was more prone to failure than in 2012. 

The structure of the paper is as follows: In Section~\ref{Sec:DefaultModel}, we describe the setup of the contagion model. 
This is followed by Section~\ref{Sec:MeanFieldSolution}, where we state our assumptions and define an iteration function that describes the contagion process. 
In Section~\ref{Sec:Results}, we discuss the results and implications of the model. 
In particular, Section~\ref{Sec:PA} addresses the stability of the system and discusses the implication for central banks and regulators to influence the quantities of a balance sheet to create, or return, to a stable system. Section~\ref{Sec:Sim} compares the mean-field results with the equilibrium values of simulation testing the robustness of the model for different random distributions for the balance sheet quantities, determining the effects of different structures for the exposure network and collateralized lending. In Section~\ref{Sec:Data}, data of US and UK banks' balance sheets are used to determine the stability of the US and UK banking system in 2007 and 2012.   
Conclusions and perspectives are given in Section~\ref{Sec:Con}.

\section{Stress Model}\label{Sec:DefaultModel}
In our model, we consider $M$ banks. To investigate the default process, we restrict our interest to the short term propagation of stress on the banking system. 
Specifically, we look at short lapses of time when banks just become unable to operate but still are not necessary defaulted, i.e. the point in time at which, according to the  Banking Act 2009 in British law, a decision about the future of an unstable bank has to be made. Thus, we distinguish between normally operating banks and distressed banks, such that the state of a bank is given by: 

\begin{equation}
S_i(t) = \left\{ 
  \begin{array}{l l}
    1 & \quad \textrm{if bank $i$ is operating normally}\\
    0 & \quad \textrm{if bank $i$ is distressed}
  \end{array} \right..
\end{equation}

\begin{figure}[h!]
\includegraphics[width=0.95\textwidth]{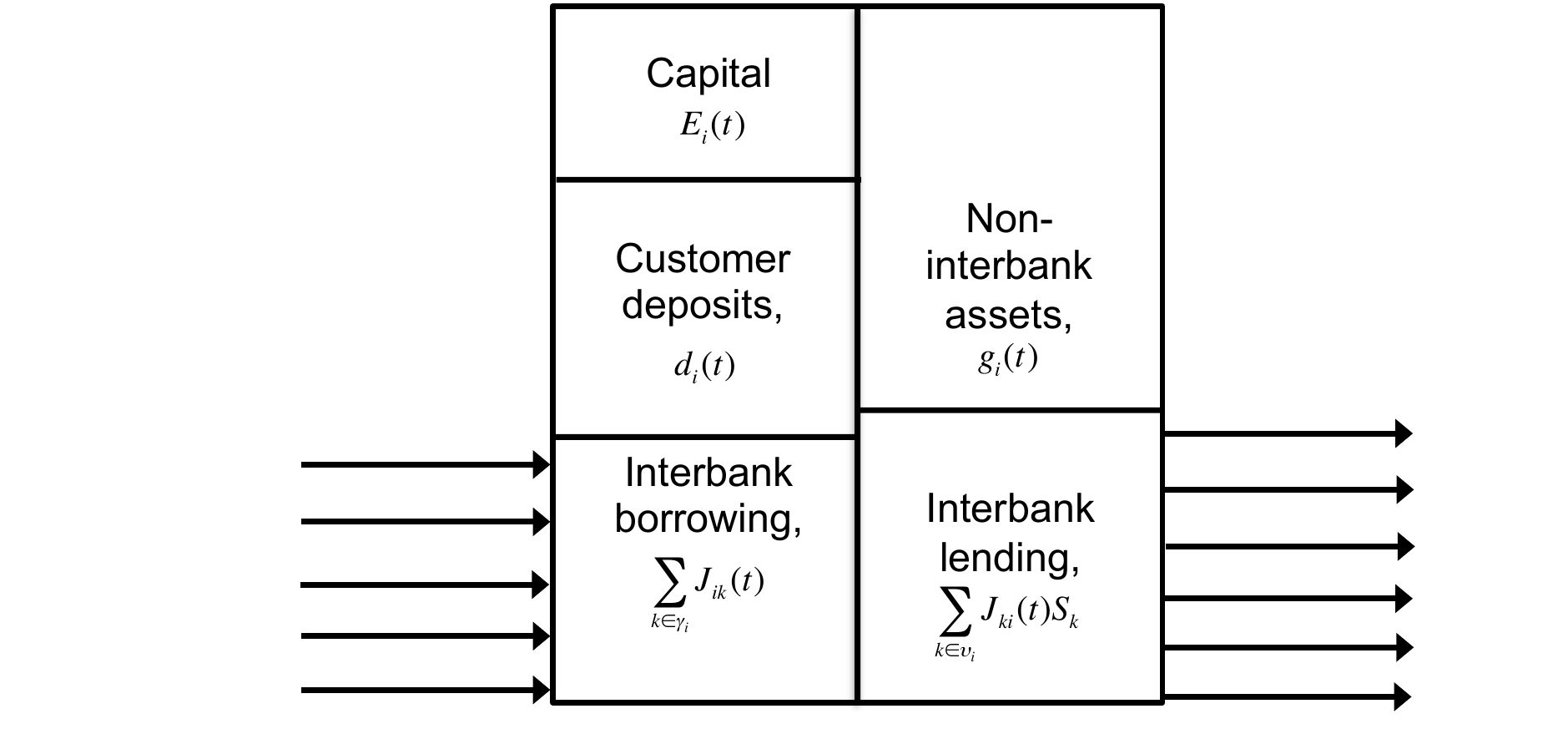}
\caption{Stylised balance sheet of bank $i$. The total liabilities of bank $i$ at time $t$, $L_i(t)$, is the sum of the bank's  deposits, $d_i(t)$, and interbank borrowing, $\sum_{k \in \gamma_i} J_{ik}$. The total assets of bank $i$ at time $t$, $A_i(t)$, is the sum of non-interbank assets, $g_i(t)$, and interbank lending, $\sum_{k \in \nu_i} J_{ki}(t)S_k(t)$. The difference in the bank's total assets and liabilities is the bank's capital $E_i(t)=A_i(t)-L_i(t)$. A bank is said to operate normally if $A_i(t)  \geq L_i(t)$. If $A_i(t) < L_i(t)$ the bank is said to be in distress.}
\label{fig:balancesheet}
\end{figure}

We adopt the stylized balance sheet introduced by \cite{Gai2007} and \cite{Nier2007}, also considering liabilities and assets. A schematic diagram of a simple balance sheet of a bank $`i$' is given in Figure~\ref{fig:balancesheet}.

The non-interbank assets of a bank $i$ at time $t$ are $g_i(t)$.  
The exposure matrix $\{J_{ki}(t)\}_{0\leq k,i \leq M}$ describes the interbank lending network at time $t$; interbank lending is modelled by adding all the debt of banks $k$ at time $t$ to a bank $i$, and multiplying this with the state of banks $k$, i.e. $\sum_{k \in \nu_i} J_{ki}(t)S_k(t)$, where $\nu_i$ is the set of borrowers to bank $i$. 
The state of a bank $k$ indicates whether bank $k$ is able to pay back the loan. 
Then, the total assets of bank $i$ at time $t$ are 

\begin{equation}\label{eqn:Assets}
A_i(t) = g_i(t) + \sum_{k \in \nu_i} J_{ki}(t)S_k(t).
\end{equation}
The customer deposits of bank $i$ at time $t$ is denoted by $d_i(t)$. 
Interbank borrowing at time $t$ is $\sum_{k \in \gamma_i} J_{ik}$, where $\gamma_i$ is the set of loaners to bank $i$ and $J_{ik}(t)$ is the amount borrowed by bank $i$ from bank $k$ so that the total liabilities of bank $i$ are:
\begin{equation}
L_i(t) = d_i(t) + \sum_{k \in \gamma_i} J_{ik}.
\end{equation}
The difference between total assets and total liabilities  of a bank $i$ is the banks capital:

\begin{equation}
E_i(t) = A_i(t)-L_i(t).
\end{equation}
The above equation is the Balance Sheet Equation. For the purpose of this model, we consider loss absorbing Tier 1 capital as capital only. 

\begin{table}[h!]
\caption{Variables of bank $i$ at time $t$ used in the balance sheet model. $S_i(t)$ describes the state of bank $i$, $J_{ik}(t)$ the loan from bank $k$ to bank $i$. Similarly, $J_{ki}(t)$ the loan from bank $i$ to bank $k$. The set $\nu_i$ denotes the set of loaners and borrowers on the interbank market of bank $i$ such that the total interbank loans from bank $i$ to bank $k$ are $\sum_{k \in \nu_i} J_{ki}(t)S_k(t)$, and bank $i$ borrows on the interbank market a total amount of $\sum_{k \in \gamma_i} J_{ik}$ from its loaner banks $k$. The non-interbank assets are represented by $g_i(t)$, and $d_i(t)$ denotes customer deposits and $E_i(t)$ denotes the bank's loss absorbing capital. The total assets are $A_i(t) = g_i(t) + \sum_{k \in \nu_i} J_{ki}(t)S_k(t)$, and the total liabilities are $L_i(t) = E_i(t)+d_i(t) + \sum_{k \in \gamma_i} J_{ik}$. }
\begin{tabular}{l  l} \hline\noalign{\smallskip}
Variable & Description of variables \\ \hline\noalign{\smallskip}
$S_i(t)$ & State \\ 
$J_{ki}(t)$ & Interbank loan from bank $k$ to bank $i$ \\ 
$g_i(t)$ & Non-interbank assets of bank $i$ \\  
$E_i(t)$ & Capital  \\  
$d_i(t)$ & Customer deposits  \\  
$\sum_{k \in \nu_i} J_{ki}(t)S_k(t)$ & Total interbank loans \\ 
$\sum_{k \in \gamma_i} J_{ik}$ & Total interbank borrowing \\ 
$A_i(t)$ & Total assets \\ 
$L_i(t)$ & Total liabilities \\ \hline\noalign{\smallskip}
\end{tabular}
\label{Tab:Variable}
\end{table}

The stress criteria is modelled by using the balance sheet test to determine insolvency, as outlined in \cite{Goode2010}. Namely, a bank is said to be in stress if assets are less than  liabilities at time $t$, i.e. the Distress Condition is:
\begin{equation}\label{Eq:StressCondition}
A_i(t) < L_i(t).
\end{equation}
Table~\ref{Tab:Variable} summarizes the variables used in this model. 

Given that the state of a bank $i$ is determined by the Distress Condition~\ref{Eq:StressCondition}, consequently the state of a bank at time $t+1$ is 

\begin{equation}\label{Eq:HeavisideFunction}
S_i(t+1) = H(A_i(t) - L_i(t)),
\end{equation}
where $H(x)$ is the Heaviside function. In the Ising model literature describing spin systems, $U_i(t) = A_i(t) - L_i(t)$ is called the `incentive function' \cite{De2006}. The probability of bank $i$ to be in a particular state, using the logit rule (which is a standard choice to determine the probability of a spin being in a particular state) is:

\begin{equation}
\begin{array}{c}
P(S_i(t) = 1|U_i(t-1))= \frac{1}{1+\exp(-\beta U_i(t-1))},
\end{array}
\end{equation}
where $\beta$ is the inverse temperature of the spin system. When $\beta$ tends to zero (infinite temperature limit) the incentive do not influence the state of the bank. 
Hence, bank $i$ is normally operating or under stress with probability $1/2$. 
Conversely, when $\beta$ tends to infinity (zero temperature limit) than Eq.~\ref{Eq:HeavisideFunction} is recovered. Thus, our stylized banking system is a zero temperature Ising model. 

\section{Uniform, Mean-Field solution}\label{Sec:MeanFieldSolution}

In order to obtain a closed form expression for the stability of the banking system, let us here introduce a few assumptions.

We are looking at the instantaneous stress imposed on a banking system given a particular distribution of non-interbank assets and liabilities. Hence, any changes in the investment after the system is distressed are neglected, as the time to counteract is considered longer than the instantaneous stress imposed by distressed banks to its creditors. Therefore, we consider most of the balance sheet quantities to be constant in time. Specifically, we consider that the process of stressing a bank, and the consequent loss of the interbank loan, are much more imminent than the distribution of any assets belonging to a distressed bank. Therefore, even if the creditor of a bank is bankrupt, the bank has still to pay any outstanding loans towards the defaulted bank, assuming further that transfers of asset belonging to the distressed bank to counter parties are excluded. Hence, we say that the liabilities, $L_i(t) = L_i$ are constant in $t$ and vary from bank to bank as drawn from a random distribution. 

The non-interbank assets $g_i(t) = g_i$ are also considered constant in $t$ and drawn from a random distribution. This represents different investment decisions, and henceforth, different investment returns. For interbank loans, we assume a mean field, i.e. the average amount loaned by bank $i$ to all its debtors, $\sum_{k \in \nu_i} J_{ki}(t)S_k(t)$,  is approximated with $z J p_t$, where $z$ is the average number of banks that are borrowing money from a given bank, $J$ is the average loan borrowed from one bank to another and $p_t$ is the fraction of  operating banks at a given time $t$. Finally, we consider that the number of banks in the system $M$ and the bank interconnections are very large. 

Let $p_{r}$ be the fraction of normally operating banks after $r$ rounds of default. 
By using the above assumptions, from Eq.\ref{Eq:HeavisideFunction}, we can write the fraction of non-defaulted banks after $r$ rounds of default as
\begin{equation}\label{Eq:Heaviside0}
\begin{array}{c}
p_{r} = \frac{1}{M}\sum_i H(g_i + z J p_{r-1} - L_i), \\ 
\end{array}
\end{equation}
which is 
\begin{equation}\label{Eq:Heaviside}
\begin{array}{c}
p_{r} =  F(p_{r-1}), \\ 
\end{array}
\end{equation}
where $F(x) = 1 - P(g_i - L_i <  - z J x)$ is a cumulative distribution function (CDF). 
Given the initial fraction, $p_0$, of surviving banks (note that $p_0$ can differ from one), the solution of Eq.\ref{Eq:Heaviside} is a fixed point probability satisfying $p=F(p)$ that may depend on the initial fraction of distressed banks. 
Note that, a distressed bank ($S_i(t)=0$) can recover and change its state to $S_i(t+1)=1$ if the difference between liabilities and total assets is positive: $A_i(t) > L_i$. 
This possibility can occur whenever capital is introduced to a distressed bank, as done via quantitative easing (QE) or government bail-outs. 
The cost of returning to a stable system, and more details about capital injections are discussed in Section~\ref{Sec:PA}.  

Let us here use the assumption, that $g_i$ and $L_i$ are independent and follow distributions in the location-scale family with mean $\mu_g$ and $\mu_L$, and standard deviation $\sigma_g$ and $\sigma_L$, respectively. 
Then the random variable $g_i-L_i$ has mean $\mu=\mu_g-\mu_L$ and  standard deviation  $\sigma=\sqrt{\sigma_g^2+\sigma_L^2}$. Thus, the mean $\mu$ can be thought of as the mean of the loss absorbing capital minus the average interbank lending and the standard deviation $\sigma$ represent the level of uncertainty of the predicted value for non-interbank assets and liabilities.  

For convenience, let us introduce the following two variables: 
\begin{equation}\label{definitionOfa}
a = \frac{\mu_{L}-\mu_{g}}{\sigma};
\end{equation}
 and 
\begin{equation}\label{definitionOfb}
b = \frac{zJ}{\sigma}.
\end{equation} 

In the following we will assume that the random variables are drawn from a normal distribution. Note, this assumption is not necessary and other kinds of distributions can be explored with a similar approach. To transform the CDF into a standard normal CDF, let $g_i -L_i = \mu + \sigma \epsilon_i$, where $\epsilon_i$ is taken from a standard normal distribution. Then the Distress Condition, Eq.~\ref{Eq:StressCondition}, becomes $\epsilon_i <a -b p_r$. 

Let us note that Eq.~\ref{Eq:Heaviside} belongs to the group of Random Field Ising models (RFIM) or moving equilibrium models in the innovation diffusion literature \cite{Bouchaud2012}. In the RFIM literature, the parameter $b$ models the influence of agents on other agents. 
In our model $b$ describes the average loan divided by a fixed variance of the non-interbank assets and liabilities, and hence, $b$ is always positive. When $b$ is negative, banks would have to pay their loaners to keep the loans. When $b$ equals zero, then, whether a bank is distressed, depends solely on the distributions of the non-interbank assets and liabilities. Since, the variables $\epsilon_i$ are from a standard normal distribution, it is to be expected that half the banks are distressed when $a$  and $b$ equal zero. Conversely, when $b$ becomes larger, i.e. when the average total interbank loan becomes larger, or when the sum of the variances on the non-interbank assets and liabilities becomes smaller, then, for a fixed $a$, the system is more resilient. However, we will see in Section~\ref{Sec:PA}, there exists a critical value $b_c $ at which the behaviour of the system changes from a smooth decline in normally operating banks to a sudden decrease. 

The parameter $a$ is the difference between the mean values of liabilities and non-interbank assets divided by the variance of non-interbank assets and liabilities. If $a$ is negative, then the mean value of non-interbank assets are larger than the liabilities. The denominator of $a$ is  the standard deviation, $\sigma$, of the sum of the variance of non-interbank assets and the liabilities. Therefore, if $a$ is negative, then a small $\sigma$ implies that $a$ becomes even more negative, leading to a more stable system. 
However, if $a$ is positive, then a small $\sigma$ leads to a more unstable system. Instead, if the mean value of non-interbank asset is sufficient to counter the liabilities, i.e. $a\ll 0$, a large $\sigma$ would imply that for some banks, their non-interbank assets would not be enough to satisfy the Distress Condition~\ref{Eq:StressCondition}. Henceforth, if the interbank loans are not sufficient, these banks are under stress. Conversely, if $a \gg 0$, then a large $\sigma$ is desirable, as this implies that for some banks, their non-interbank asset value is higher than the average value. Thus, these banks can satisfy the Distress Condition~\ref{Eq:StressCondition}, and will operate normally.   

\section{Results}\label{Sec:Results}

\subsection{Fixed points}\label{Sec:FP}
To study the behaviour of the iteration map in Eq.~\ref{Eq:Heaviside}, we investigate when it reaches a fixed point $p$, such that $p=F(p)$. 
From Eq.~\ref{Eq:Heaviside}, by using the standard normal distribution, we can write:
\begin{equation}\label{MeanFieldSurv}
F(x) = 1 - \Phi(a - bx),
\end{equation}
where $\Phi(x)$ is the standard normal CDF for $x \in [0,1]$. 
Note, that the following discussion can be repeated with another location-scale distribution.  

\begin{figure}[h!]

\includegraphics[width=0.95\textwidth]{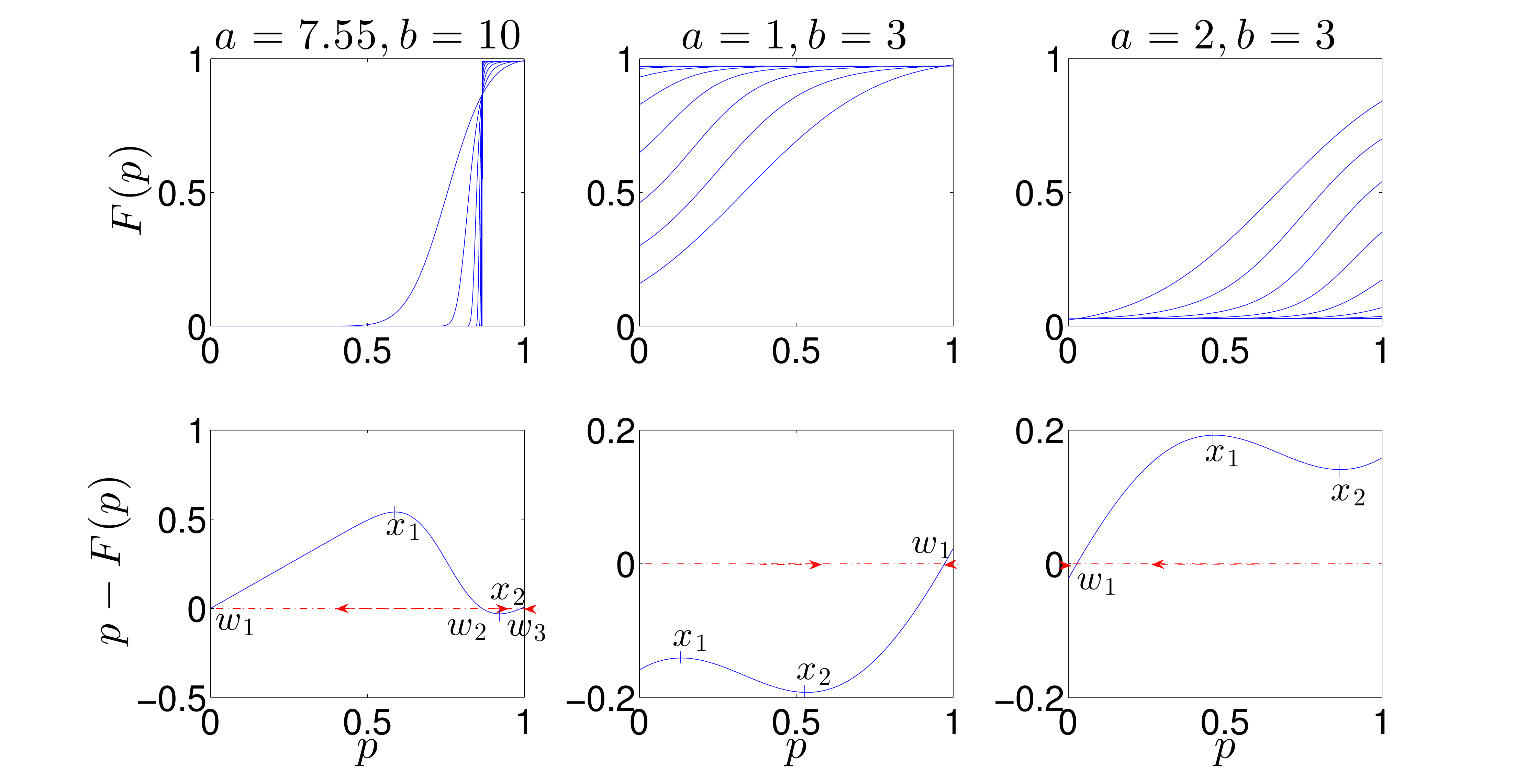}
\caption{
The first row of this figure shows $F(p)$ form Eq.~\ref{MeanFieldSurv} vs. $p$ for $r=0,...,100$ with various combinations of parameters $a$ and $b$.
The second row shows plots of $p-F(p)$. The extreme values, $x_1$ and $x_2$, of $p-F(p)$ are indicated with a cross and the corresponding fixed points, $w_1, w_2$ and $w_3$ are the points where $p-F(p)$ crosses zero. The arrows indicate which fixed point is reached starting at a particular $p_0$.
}
\label{fig:fixedpointeuquationanditerationfunction}
\end{figure} 

In order to investigate the fixed point we report in Figure~\ref{fig:fixedpointeuquationanditerationfunction} (first row) various plots of the iteration map $F(x)$ for different values of $a$ and $b$ for $r$ from $r=0$ to $r = 100$. 
It becomes clear, that, given particular parameter values, and the same starting value, the fixed points change. 
This is better illustrated in the second row of Figure~\ref{fig:fixedpointeuquationanditerationfunction} where $p-F(p)$ is plotted which crosses zero at the fixed point. It becomes clear, that up to three fixed points can occur. 
Namely, if $b< b_c = \sqrt{2 \pi}$, only one fixed point occurs.
If $b>b_c$, then three fixed points,  $w_1,w_2,w_3$, become possible. That is because, whenever $b >  b_c$, the  function $x-F(x)$ has extrema at $x_{1,2} = b^{-1}(a\mp\sqrt{2 \ln\frac{b}{ \sqrt{b_c}}})$  (where $x_1$ is a maximum and $x_2$ is a minima)  
if $a \in [a_1, a_2]$, where $a_1 = b + \sqrt{2\ln\frac{b}{b_c}}-b\Phi(\sqrt{2\ln\frac{b}{b_c}})$ and $a_2 = b - \sqrt{2\ln\frac{b}{b_c}}-b\Phi(-\sqrt{2\ln\frac{b}{b_c}})$. 
Note that, $w_1 \leq x_1 \leq w_2 \leq x_2 \leq w_3$. 
We have $F'(w_1) < 1$ indicating that the fixed point $w_1$ is stable. 
Similarly,  $w_3$ is stable, and, because the iteration map is one-dimensional, $w_2$ is unstable. 
Thus, the fixed point $w_2$ forms a barrier. 

If $a<a_1$, or $a>a_2$, then $p=F(p)$ has only one solution, $w_1$, which is a stable fixed point.

Consider the case when $a\in (a_1, a_2)$. If the starting value $p_0$ is in the orbit $[0,w_1]$ or $[w_3,1]$, then the attracting fixed points are $w_1$  or $w_3$, respectively. 
If $p_0 \in [w_1,w_2]$, then $w_2$ is a repelling fixed point and $w_1$ is the attracting fixed point that is eventually reached. Similarly, if $p_0 \in [w_2,w_3]$, the fixed point eventually reached is $w_3$.  

If $a = a_1$, then the fixed points $w_1$ and $w_2$ merge, and $w_1=x_1=w_2$. This implies that the left-hand side of the fixed point $w_1=w_2$ is stable, however the right-hand side of the fixed point $w_1=w_2$ is unstable. Hence, if a starting value $p_0$ is in the orbit $[0,w_1=w_2]$, then the fixed point reached is $w_1$. However, if $p_0 \in [w_1=w_2, w_3]$, then the fixed point reached is $w_3$. For $p_0 \in [w_3, 1]$ the attracting fixed point is again $w_3$. 

In the case if $a = a_2$, then $w_2 = x_2 = w_3$, i.e. $w_2$ and $w_3$ merge implying that if $p_0 \in [w_1,w_2=w_3]$, then $w_1$ is the attracting fixed point. 
If $p_0 \in [0,w_1]$ or $p_0 \in [w_3, 1]$, then the fixed points reached are $w_1$ and $w_3$, respectively.  

In terms of the stability of the modeled banking system we note that when $b> b_c$ a barrier, represented by the unstable fixed point, can occur, such that the number of operating banks does not decrease below a certain value (or increases above a certain value). However, if there is a  change in the parameter values, then it becomes possible that the entire system suddenly collapses (or becomes fully functional again). Hence, for $b<b_c$, the system is reversible, but for $b>b_c$, a hysteresis cycle occurs, such that the system becomes irreversible, and depends on its history.  
Therefore, a large amount of lending on the interbank market (i.e. large $b$ when $p_0 = 1$) can help to stabilize the system, if the corresponding value for liabilities and mean value of non-interbank assets are such that $a < a_2$, because in this case the barrier prevents an entire system failure. 

\subsection{Change in the number of surviving banks induced by one bank failure}\label{Sec:BF}

For a small change from $p_r$ to $p_{r+1}$, the change in the number of surviving banks is given by $M F'(p_r)$ . 
Note that $F'(x)$ is the probability density function that $\epsilon_i = a-bx$. 
Thus, the number of banks becoming distressed as a consequence of one bank changing from operating normally to distressed in the next iteration is \cite{Dahmen1996}:

\begin{equation}
n = F'(x).
\end{equation}   
If $n$ is less than one, any avalanche will eventually stop. Whereas, if $n \geq 1$  one bank default can trigger an entire system failure. Starting with $p_0 =1$, for $b>b_c$ and $x = x_1$, $n$ is precisely one. 
The maximum of $F'(x)$ is reached when $x = a/b$. 
At this point the number of stressed banks triggered by one bank in the following iteration is of order $z$ suggesting that all the neighbouring banks of the initially distressed bank become all distressed as well.  

\subsection{Relation between $a$ and $b$}

For fixed capital $E(p)$, the parameters $a$ and $b$ are dependent on one other such that the parameter $a$ can be expressed in terms of $b$ as:

\begin{equation}\label{Eq:Relationshipab}
a = -\frac{E(p)}{\sigma} +bp_r.
\end{equation}
where $E(p) = \mu_g + bp \sigma -\mu_L$. Thus, a change in $a$ given a fixed $b$ at the fixed point $p$ can only happen when external capital is introduced to the system. There are multiple ways of increasing capital of a bank. For instance, a bank can raise capital by issuing shares. Given the thread of default a government can intervene by inducing capital into the distressed bank via government bailouts. Further, central banks use methods of QE by adjusting interest rates and lending to banks, or buying assets using open market operations. Hence, QE can ensure that liabilities are reduced using the central bank loans with smaller interest rates than otherwise required by the interbank market and assets are liquidated above the market value ensuring that capital is not needed to overcome losses when faced by liquidity shortages.

\subsection{Parameter analysis}\label{Sec:PA}  

\begin{figure}[h!]

\includegraphics[width=0.95\textwidth]{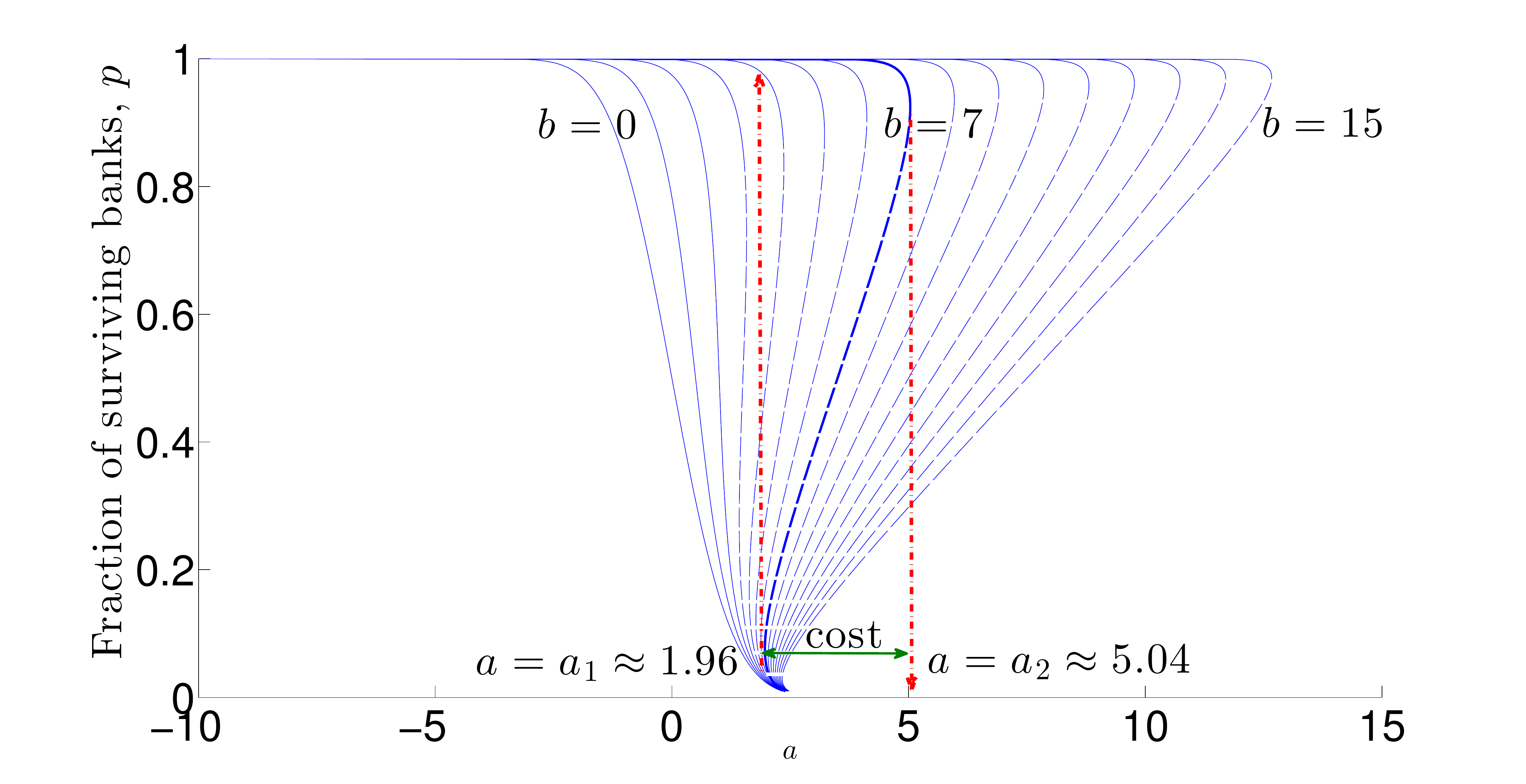}
\caption{This figure shows the fraction of surviving banks as a function of the parameter $a$ for given fixed values of $b$. The blue graphs are the solution of the Iteration Function~\ref{MeanFieldSurv} for different $b$ values whereby $b = 1,...,15$. The solid lines indicate stable fixed points whereas the dotted lines indicate unstable fixed points. If the fixed point is unique as in the case for $b = 1,2$, no hysteresis occurs for decreasing or increasing $a$.  For this value of $b> b_c$ and a particular range of $a$ three fixed points become possible leading to a hysteresis cycle. The thick blue line indicates the fixed points for $b = 7$. The red arrows indicate the hysteresis cycle that occurs for $b = 7$. Starting from $p_0 = 1$, the parameter $a$ needs to increase to $a=a_2 \approx 5.04$ for the entire system to default. If the starting value is $p_0 = 0$ then $a$ needs to decrease to $a = a_1 \approx 1.96$ for the banks to be operating. Thus, the path is history depended. }
\label{fig:iterationfunction2}
\end{figure}

We have observed that when $b$ becomes larger than the critical value $b_c$, the system passes form a reversible kind of dynamics to an irreversible one where hysteresis cycles emerge. 
This is illustrated in Figure~\ref{fig:iterationfunction2} where the fixed point probability values are plotted for varying $a$ for various $b$ ranging from $b = 0,...,15$. The solid blue lines indicate the stable fixed points, whereas the blue dashed lines indicate the unstable fixed points. The hysteresis cycle is indicated by the red arrows.

We can observe that at $b = 0$, when banks are not lending to each other, the system is stable for negative values of $a$; fluctuations in the  assets side of the balance sheet equation can cause banks to fail and, at $a=0$, half the banks in the system are in distress. By lending money from one bank to another ($b>0$), the system becomes more stable with smaller numbers of banks in distress for the same values of $a$. 

If $a$ increases further but $b$ is kept constant, then more banks fail as the difference between the banks non-interbank assets and liabilities increases. Hence, the capital in the system is lowered. If $b$ is below its critical value, then the system is reversible and all fixed points are stable. If $b$ becomes larger than the critical value $b_c$ and $a<a_2$, almost the entire system is stable (if $p_0 = 1$) because of the barrier. 
However, when $a$ increases above $a_2$, then the whole system suddenly crashes.

Also, if $a$ is constant but $b$  decreases, then a sudden jump becomes possible as well. 
Let us here note that a decrease in $b$ happens, if the average interbank loans $zJ$ decrease, or the variance $\sigma= \sqrt{\sigma_g(t)^2+\sigma_L(t)^2}$ increases. In \cite{Iori2008}, it was shown that during the financial crisis, there was indeed a decrease in the amount of money loaned but also the interest rates for loans increased. Thus, $b$ decreased, and $a$ increased. 
In our stylized system this is a mechanism that would create disastrous consequences unless $b<b_c$.

In order to return to a normally operating system after the crash, $a$ needs to be reduced at least to $a_1$. 
Then a sudden jump brings the whole system operative again. 
Hence, the cost of rescuing a banking system is given by the difference between $a_1(b(t))$ and $a_2(b(t+\delta t))$, where $b(t)$ is the value of $b$ at the beginning of the crisis and $b(t+\delta t)$ the value of $b$ at the time of rescue.  

To be more specific, let us here discuss the case $b =7$ and starting from fully operating banks (i.e. $p_0 = 1$). 
Here the infinite avalanche occurs when $a$ reaches $a_2\approx 5.04$. 
Whereas, if one  starts with all banks distressed, $a$ would need to be lowered to $a_1 \approx 1.96$, in order to return to a stable system. In Figure~\ref{fig:iterationfunction2}, this cost is indicated by the green arrow.

\begin{figure}[h!]
\includegraphics[width=0.95\textwidth]{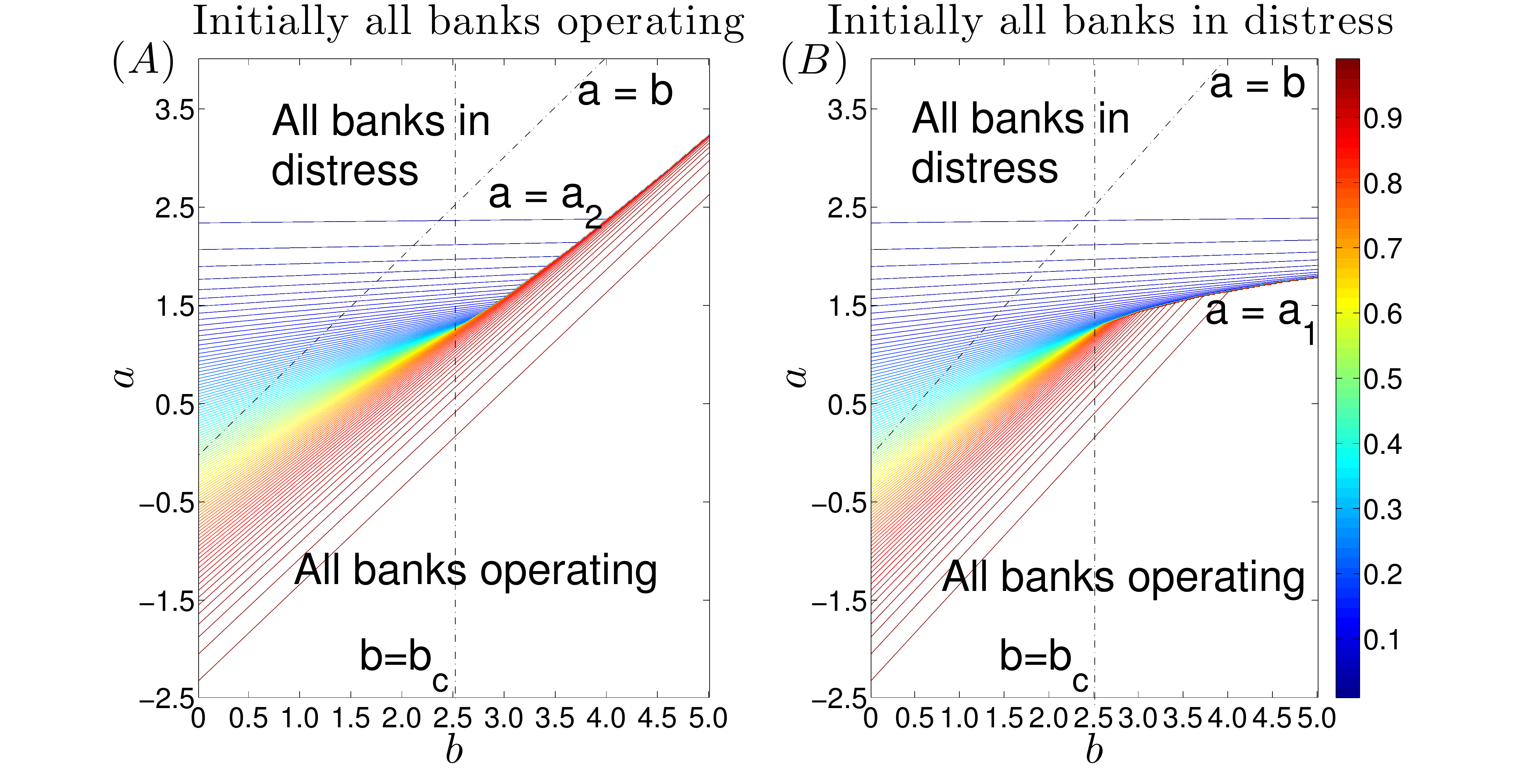}
\caption{The figures show the fraction of operating banks for given $a$ and $b$ obtained by numerically solving the Iteration Function~\ref{MeanFieldSurv} starting from an initial value $p_0 =1$  (plot $A$) and $p_0 = 0$ (plot $B$). The hysteresis behaviour becomes visible form the jump occurring in ($A$) at $a = a_2$ and in ($B$) at $a = a_1$.}
\label{fig:phasespace3}
\end{figure}

Figure~\ref{fig:phasespace3} is a plot of the equilibrium fraction of normally operating banks for different parameter values. The figure contains two plots $A$ and $B$, and depicts the solution of Eq.~\ref{Eq:Heaviside} for different values of $a$ and $b$ when the initial state of all banks is $p_0 = 1$ (plot $A$) or $p_0 = 0$ (plot $B$). Whenever $b=0$, the fraction of operating banks depends only on the CDF of non-interbank assets. In the case, of the standard normal CDF, for $a=0$ half of the banks are expected to be under stress; at $a = -2.5$ the equilibrium fraction of operating banks is $p \approx 0.9938$; whereas for $a = 2.5$ the equilibrium fraction of operating banks is $p \approx 0.0062$. 
If $0<b<b_c$, the  system becomes more stable which is obvious as the asset side of the balance sheet is increased and the interbank loans act as an extra asset. If $a$ is kept constant then either extra capital is introduced in the system or the values $\mu_L, \mu_g, \sigma_L$ and $\sigma_g$ change such that $a$ stays constant. Further, for values of $b$ in that range, the decline in the fraction of normally operating banks for increasing $a$ is still smooth. 
When $b> b_c$, the fraction of operating banks suddenly jumps from almost all operating to almost all banks in stress, which happens because of the occurrence of the multiple fixed points as outlined in Section~\ref{Sec:FP}.

\subsection{Leverage}

\begin{figure}[h!]
\includegraphics[width=0.95\textwidth]{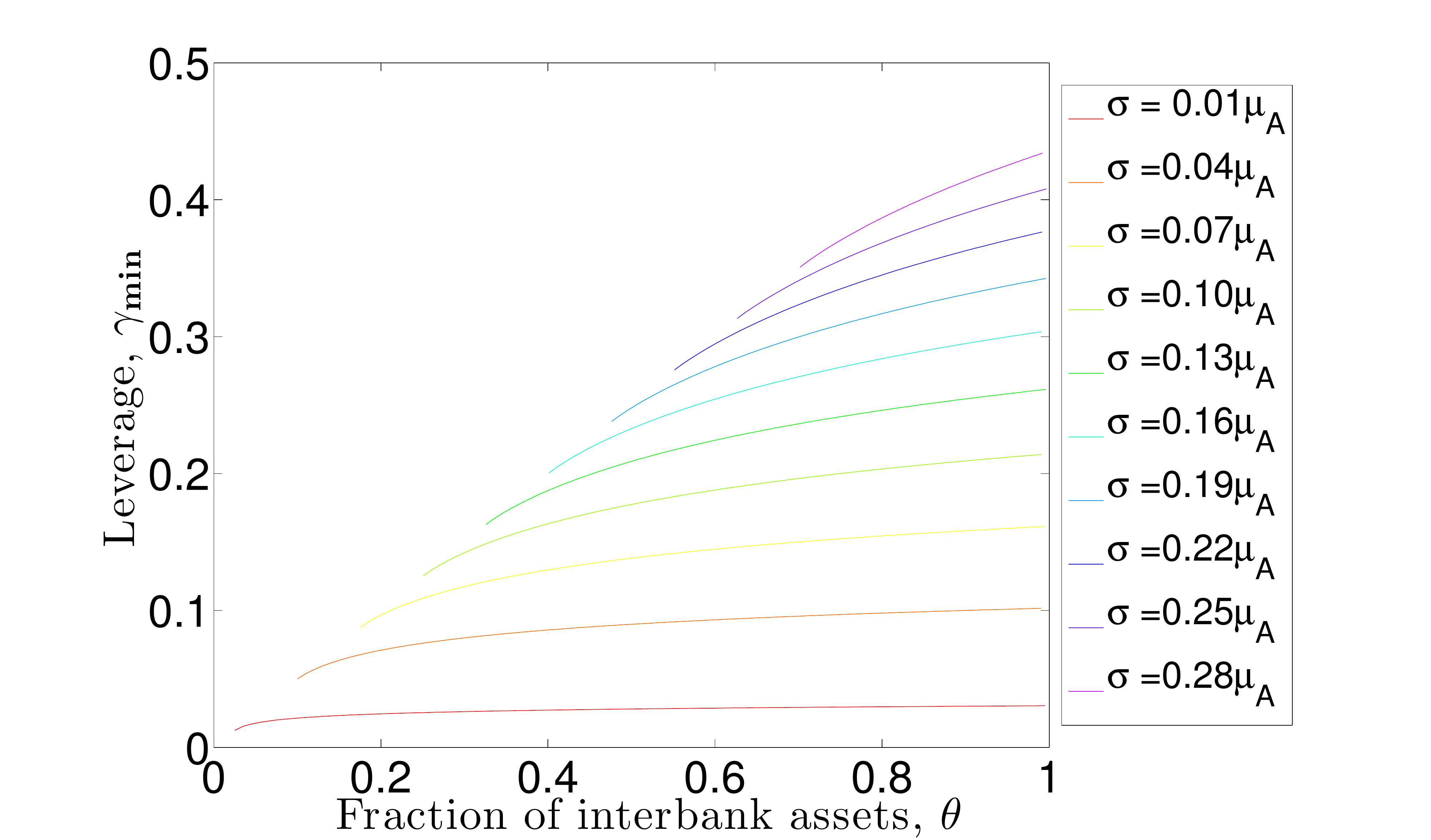}
\caption{The figure shows the minimum leverage, $\gamma_{\min}$, for an average bank to ensure a stable banking system as a function of the fraction of innterbank assets $\theta$. The different curves correspond to various $\sigma$'s. From the figure, we can conduct that the larger $\sigma$ the larger a $\theta$ is required for the jump to occur. }
\label{fig:leverage}
\end{figure}

For a stable system (i.e. $p \approx 1$) with $b>b_c$, the ratios between assets and liabilities should ensure that  $a \leq a_2$. Using the mean-field assumption, the interbank assets of a bank are a fraction $\theta$ of the total mean assets, i.e. $zJ = \theta \mu_A$ (where $\mu_A = \mu_g + zJp_0$). Further using Eq.~\ref{Eq:Relationshipab}, this implies that the leverage ratio - the ratio of capital to total assets, i.e. $\gamma = \frac{\mu_E}{\mu_A}$, should satisfy the following condition to ensure a stable banking system:
 
\begin{equation}\label{Eq:leverage}
\gamma \geq \frac{\theta_c}{b_c} \sqrt{2\ln\frac{\theta}{\theta_c}}+\theta\Phi\bigg(-\sqrt{2\ln\frac{\theta}{\theta_c}}\bigg),
\end{equation}

where $\theta_c = \frac{\sigma b_c}{\mu_A}$. Figure~\ref{fig:leverage} is a plot of Eq.~\ref{Eq:leverage} depicting the minimal value of the leverage, $\gamma_{\min}$, at which the system is stable as a function of $\theta$, of interbank assets to total assets for given values of $\sigma$. The value of $\sigma$ is chosen to be a fraction of the mean total assets for each graph as applicable in the accompanying legend. We chose to represent $\sigma$ in this way as then Eq.~\ref{Eq:leverage} becomes independent of $\mu_A$. Any leverage value above and including $\gamma_{\min}$ ensures a safe banking system given a particular $\sigma$. 

It becomes clear that the larger $\sigma$ the larger $\theta$ has to be to be greater than $\theta_c$ in order to observe the jump and henceforth a system-wide failure of the banking system. However, the leverage requirement also needs to increase significantly in order for the banking system to be stable. 

\subsection{Collateralized lending}

The effects of collateralized lending can be discussed by adding $qzJ(1-p_t)$ to the sum of total assets. The parameter $q \in[0,1]$ indicates the average amount a bank can expect as collateral when counter parties become unable to pay back loans. Another way of thinking about this term is the value of a defaulted loan, i.e. any possible payback during the insolvency procedure. The term $qzJ(1-p_t)$ shifts the point of systemic failure and the variables $a$ and $b$ including collateralized loans need to be adjusted to

\begin{equation}\label{definitionOfaa}
a' = \frac{\mu_{L}-\mu_{g}+qzJ}{\sigma};
\end{equation}
 and 
\begin{equation}\label{definitionOfbb}
b' = \frac{zJ(1-q)}{\sigma}.
\end{equation} 

A plot of the fixed points of the Iteration Function~\ref{MeanFieldSurv} using $a'$ and $b'$ is given in Figure~\ref{fig:collateral} in Section~\ref{Sec:DerivativesandCollateral}.
\section{Simulation}\label{Sec:Sim}  
     
\subsection{Simulation Set-Up}\label{Sec:SimulationSetUp} 
  
The mean-field assumption of the interbank market implies that each bank lends the same amount  to all other banks. 
Hence when using the mean-field assumption the network structure is a fully connected graph. Further, normal fluctuations for the value of the non-interbank assets have been assumed  in the analysis done so far. To test the effects of different network structures and distributions, we use a simulation approach similar to the one in \cite{Furfine2003}. The banking system modelled in our simulations still represents a highly stylized banking system as we restrict the simulations to banks with balance sheets of similar size  leaving the effects of a heterogeneous banking system to a later stage. The simulation results are intended to verify that the overall behaviour and the hysteresis effect can be retrieved also by using different network structures and different distributions for liabilities and assets.  

As before, the system consists of $M$ banks. Each bank $i$ is initially calibrated with  liabilities $L_i(0)$ and assets $A_i(0)$ drawn respectively from distributions with mean $\mu_L$ and variance $\sigma_L^2$ (for liabilities) and mean $\mu_A$ and variance $\sigma_A^2$ (for assets). For the interbank assets, we use a fraction $\theta \in [0,1]$ of the total assets $A_i(0)$ of bank $i$ such that the total interbank assets of bank $i$ are $\theta A_i(0)$. The interbank lending structure is given by the network $G= \{g_{1\leq ij \leq M}\}$, where $g_{ij} =1$ if bank $i$ loans to bank $j$ and 0 otherwise. 
The individual loans from bank $i$ to its neighbouring banks $j$ are the total interbank assets divided by the degree $z_i$ of bank $i$, i.e. the amount loaned from bank $i$ to bank $j$ is $\theta A_{i}g_{i,j}/z_i$.

The distributions tested are Normal distribution and Student's t-distribution. To calibrate total assets and total liabilities with Normal distributions, random variables $\epsilon_i$ are drawn from a standard normal distribution;  the total assets are $A_i(0)= \mu_A +\sigma_A\epsilon^A_i$ and the total liabilities are $L_i(0)= \mu_L +\sigma_L\epsilon^L_i$. Similarly, if the distribution used to calibrate total assets and total liabilities is the Student's t distribution, random variables $t^L_i$, $t^A_i$ are drawn from a standard Student's t distribution with degree of freedom $\nu$. The total assets and liabilities are given by $A_i(0)= \mu_A +\sigma_A t^A_i$ and $L_i(0)= \mu_L +\sigma_L t^L_i$. Note that the random variables $\epsilon^{L,A}_i$ and $t^{L,A}_i$ are different and independent. 

For constructing the underlying exposure network structure $G$, we used three different standard network types: the Erd\H{o}s-R\'{e}ny network, the Small-World network \cite{Watts1998} and a core-periphery network produced using the preferential attachment algorithm as outlined in \cite{Barabasi1999}. For the Erd\H{o}s-R\'{e}ny network, a bank $i$ is connected to a bank $j$ with probability $\alpha$. For the Small-World network, we used an initial network where each bank is connected to its $c$ closest neighbours and a probability $\beta$ is used to re-wire any existing links between the neighbouring banks to other banks creating the small-world effect. 
For the core-periphery network, we used an Erd\H{o}s-R\'{e}ny seed network of  banks with a connection probability of $\alpha$ and added `perioheral' banks individually to the system  using preferential attachment.    

We would like to stress that the network structures as well as the distributions are standard choices and reality might differ greatly. The  different structures and distributions are intended to show that the model predictions are robust for a variety of assumptions. The choice to use Normal and Student's t distributions is to compare the results drawn from the iteration function, as these distributions are part of the location-scale family, and the mean and variance values for total liabilities and assets can be compared to the fixed points of the Iteration Function~\ref{MeanFieldSurv}. 
The parameter values used to initialize the model can be found in Table~\ref{Tab:ModelParameter}.

\begin{table}[H]\caption{\scriptsize Variables and values used for initializing banks balance sheets and exposure structure in the simulation modelling a stylized banking system. The banking system consist of $M = 500$ banks. The state of each bank is set to operating initially, i.e. $S_i(0) =1$ for all banks $i$. Two location scale distribution, the normal distribution and the Student t distribution, are used to calibrate the balance sheets of banks. In particular, the initial value for total asset and liabilities for bank $i$ are $A_i(0)= \mu_A +\sigma_A\epsilon_i^A$ and $L_i(0)= \mu_L +\sigma_L\epsilon_i^L$ for simulations using normal distributions, and $A_i(0)= \mu_A +\sigma_A t_i^Â$ and $L_i(0)= \mu_L +\sigma_L t_i^L$ for simulations using Student t distributions. To compute the structure of the exposure network $G = \{g_{1 \leq i,j \leq M}\}$, three different network structures are used: Erd\H{o}s-R\'{e}ny networks, Small-World networks and a network structure with a core and periphery banks. For the Erd\H{o}s-R\'{e}ny networks a link exists between two banks with probability $\alpha = 0.1$. To construct the Small-World network, we used the algorithm from \cite{Watts1998} with banks have $c = 4$ neighbours and a re-wiring probability of each link of $\beta = 0.1$. To create the core-periphery network, we use the algorithm from \cite{Barabasi1999} with an Erd\H{o}s-R\'{e}ny network seed network with $50$ banks and connection probability $\alpha =0.1$ and 450 banks with 15 links added with a preferential attachment to the existing banks as described in \cite{Barabasi1999}. The weight for a loan from bank $i$ to bank $j$ is $\theta A_{i}g_{i,j}/z_i$, where $\theta$ is the fraction of interbank assets total assets.}
\small\addtolength{\tabcolsep}{-5pt}
\begin{tabular}{ l  l  l }\hline\noalign{\smallskip}
Variable & Values used for calibration & Description of variables \\ & & of bank $i$ at time $0$ \\ \hline\noalign{\smallskip}
$M$ & 500 & Number of banks in \\ & & the stylised banking system. \\ 
$\epsilon^{A,L}_i$ & $\epsilon^{A,L}_i \sim N(0,1)$ & Standard normal random variables. \\ 
$t^{A,L}_i$ & $t^{A,L}_i \sim T(\nu)$ & Standard Student's t random \\ &  & variables with degree of freedom $\nu$. \\ 
$\nu$ & 2 & Degree of freedom for \\ & & Student's t distribution. \\ 
$S_i(t)$ & $S_i(0)=1$ & State, all banks \\ & & are operating initially. \\ 
$\mu_A$ & 1000  & Mean value for \\ & & total assets.\\ 
$\sigma_A$&  30 & Standard deviation \\ & & for assets.  \\ 
$\mu_L$ & 700 - 1200  & Mean value for liabilities. \\ 
$\sigma_L$&  50 & Standard deviation \\ & & for liabilities.  \\ 
$\theta$&  0.0, 0.1, 0.3 & Fraction for  interbank assets  \\ 
&  &  Probability of bank $i$ \\ & & being connected with bank $j$, \\$\alpha$ & 0.1 & used to generate Erd\H{o}s-R\'{e}ny network \\  & & and seed network for the \\ & & core-periphery network. \\ 
$c$ & 4 & Neighbouring banks of all bank $i$ \\ & & in Small-World network. \\ 
$\beta$ & 0.1 & Re-wiring probability for a link \\ & & in the Small-World network .\\ \hline\noalign{\smallskip}
\end{tabular}
\label{Tab:ModelParameter}
\end{table}

For the contagion propagation, we use a similar algorithm as in \cite{Furfine2003}. 
Specifically, for each iteration $r$, the following algorithm is computed:

\begin{enumerate}
\item It is simultaneously tested for all banks $i$ whether the total assets of each bank $i$ is smaller than its total liabilities.
\item If this is the case, then the state of bank $i$, $S_i(r)$, is set to zero, and the bank is said to be distressed.
\item Eq.~\ref{eqn:Assets} is then used to evaluate the total assets of bank $i$ for the next iteration $r+1$.
\item The above steps are repeated until no further default occur. 
\end{enumerate}
When the iteration process stops we obtain the fraction of surviving banks $p$ by counting the banks that are still operating.

\subsection{Comparing with Fixed Point Solution}\label{Sec:ComparingFPS}

To compare the fraction of surviving banks with the fixed points of the Iteration Function~\ref{MeanFieldSurv}, we identify $(\mu_L-\mu_A)/(\sigma_A^2+\sigma_L^2)^{1/2}$ with $a-b$ and $Jz \approx \theta A_i(0)$, where $a$ and $b$ from Eqns.~\ref{definitionOfa}~and~\ref{definitionOfb}.

\begin{figure}[h!]
\includegraphics[width=0.95\textwidth]{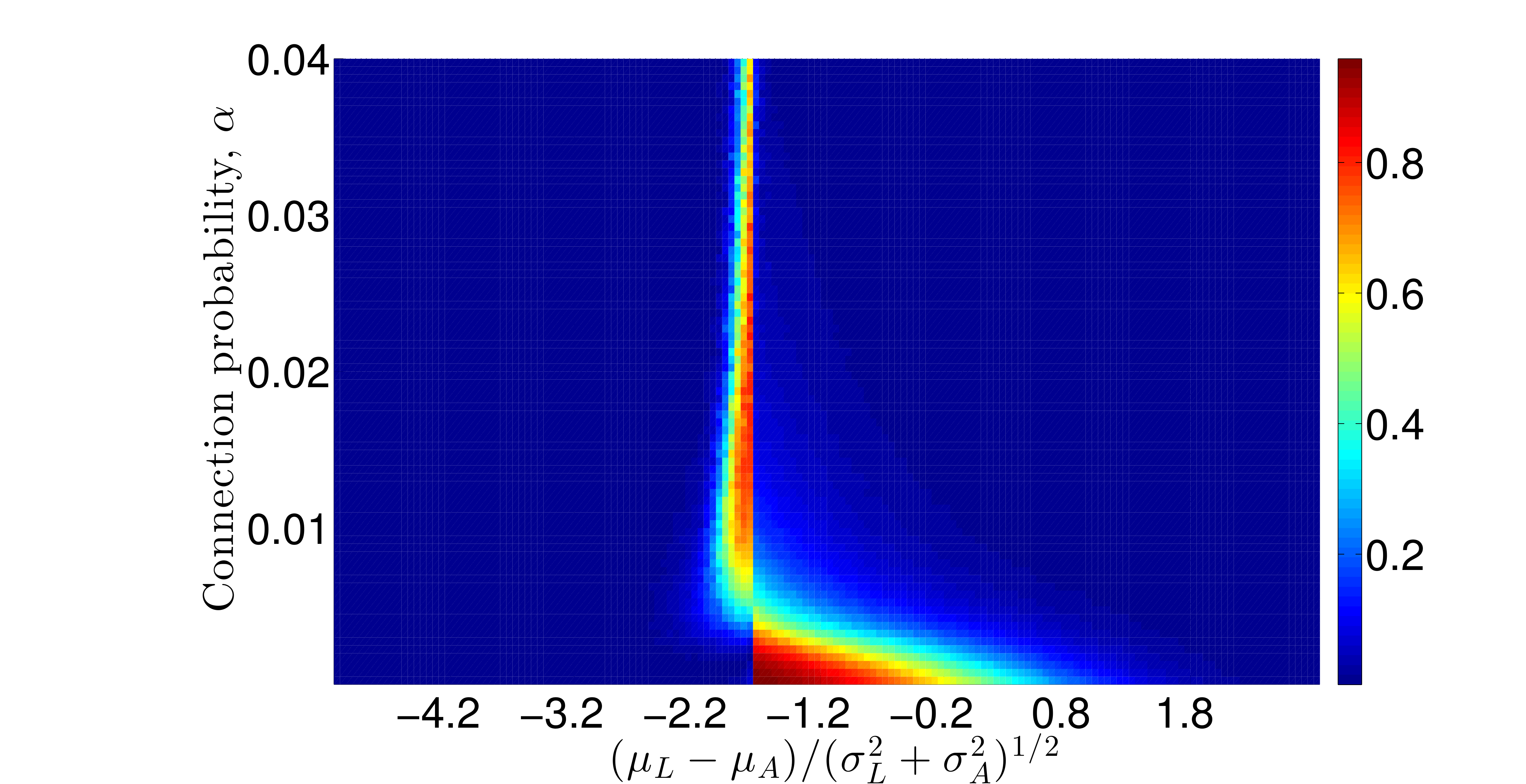}
\caption{The figure shows the average error between the solution of the simulation and Iteration Function~\ref{MeanFieldSurv} of the fraction of surviving banks. 
The figure reports the second norms of the difference between the fractions of surviving banks of the fixed point solutions of Iteration Function~\ref{MeanFieldSurv} and the  fraction of surviving banks of an average of 100 simulations for fixed values $(\mu_L-\mu_A)/(\sigma_A^2+\sigma_L^2)^{1/2}$ (changing $\mu_L$ for different simulations) and $a-b$ (changing $a$ for  different fixed points). 
The simulation assumes Normal distributions for the balance sheet values and for the structure of the exposure network Erd\H{o}s-R\'{e}ny networks with connection probability $\alpha$ and fraction of interbank loans to total assets $\theta$ are used. 
To test the influence of the number of links from one bank to others, $\alpha$ is varied in $(0,0.1]$. }
\label{fig:DegreeDifference}
\end{figure}

Figure~\ref{fig:DegreeDifference} shows the difference between the fraction of surviving banks computed by using the fixed points of the Iteration Function~\ref{MeanFieldSurv} and the mean value of the fraction of surviving banks from 100 simulations. In the simulation, we use Erd\H{o}s-R\'{e}ny networks as underlying structures for the exposure networks and Normal distributions for liabilities and assets with varying mean of the Liabilities $\mu_L$ and connection probability $\alpha$. The ratio between interbank assets and total asset $\theta$ is set to 0.3. For this value of $\theta$, $b$ is well above its critical value and a jump is predicted. For the fixed point equation $a$ is varied to balance the changes in $\mu_L$ in the simulation. The colour scale in Figure~\ref{fig:DegreeDifference} reports  the error between the predicted values and the value archived using the average from 100 simulations. 
As expected close to the jump the error is large. However, also for $\alpha$ smaller than 0.03, a large error is observed. 
This is because in that region the jump is only marginal or does not occur in the simulation implying that due to the smaller number of links the stress distribution and subsequent cumulative counter party losses via the network are not realized.

We note that large errors happen in a range close to the jump for connection probabilities $\alpha$ smaller than $8\cdot	10^{-3}$. 
In that region the average degree $\bar{z}$ of a bank is between $0$ and $4$ for $M=500$\footnote{For an Erd\H{o}s-R\'{e}ny network is the average degree is $\bar{z}=\alpha (M-1)$}. 
For $\alpha < 10^{-3}$, the jump was not observed or it was not very dominant in the simulation testing. 
The amount loaned from one bank to another is still $\theta A_i$. However, it is a well known phenomena that the upper critical Euclidean dimension for the mean-field assumption of the Ising model is 4 \cite{Barrat2008}. Thus, it becomes clear that the mean-field approximation does not capture the behaviour for average degrees smaller than 4 and further investigation needs to be done into whether an average low number of counter parties in a banking system reduces the risk of a systemic stress event. 

\subsection{Normal and Student's t Distributions}\label{Sec:NormalStudent}

\begin{figure}[h!]
\includegraphics[width=0.95\textwidth]{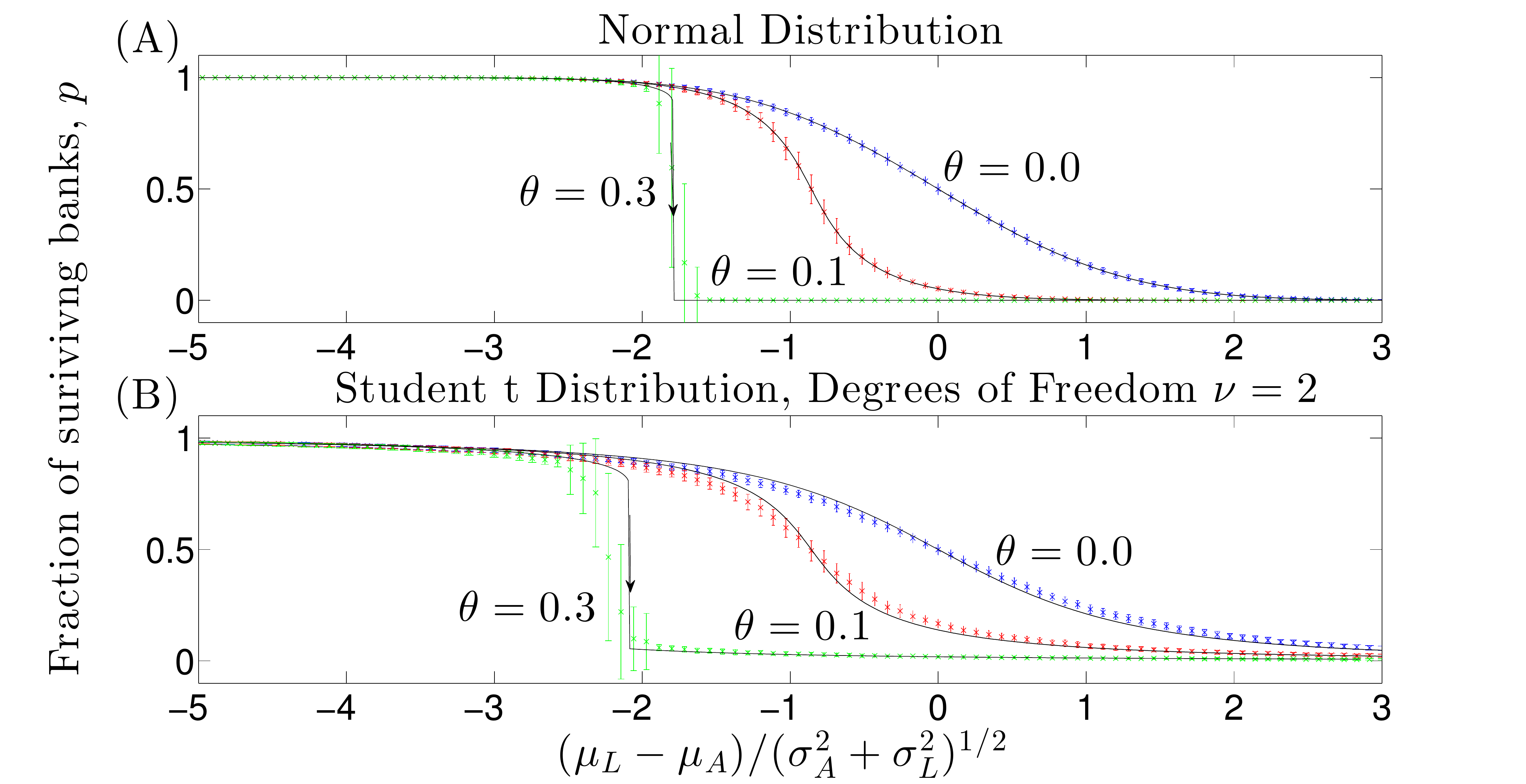}
\caption{The figure shows the fraction of surviving banks $p$ evaluated by initializing the liabilities and assets of banks' balance sheets with Normal distributions ($A$) and Students't distributions with 2 degrees of freedom ($B$) with varying mean $\mu_L$ and fixed standard deviation $\sigma_L$ for liabilities, fixed mean $\mu_A$ and fixed standard deviation $\sigma_A$ for assets plotted against $(\mu_L-\mu_A)/(\sigma_A^2+\sigma_L^2)^{1/2}$. Each symbol is the average of the fraction of surviving banks of 100 simulations. The error bars are the standard deviation of the 100 simulations. To compute the blue line, we set the average fraction of interbank loans to zero, i.e. $\theta = 0.0$, for the red line $\theta$ was set to $0.1$ and for the green line $\theta$ was set to 0.3. The underlying structure of the exposure networks are Erd\H{o}s-R\'{e}ny networks with connection probability $\alpha = 0.1$ and $M = 500$ banks. The black lines accompanying each plot are the fixed points of the Iteration Function~\ref{MeanFieldSurv} plotted against $a-b$ which is approximately $(\mu_L-\mu_A)/(\sigma_A^2+\sigma_L^2)^{1/2}$. Note that $b$ is changed to fit the equivalent $\theta$ value. A steep decline in the fraction of surviving banks happens when $\theta$ equals to 0.3 in the area of the predicted jump. For $\theta$ equal to 0.0 and 0.1 the simulation result for both distributions are close to the fixed point solution of the Iteration Function~\ref{MeanFieldSurv}. The parameter values used to initialize the system are stated in Table~\ref{Tab:ModelParameter}.}
\label{fig:DifferentDistributions}
\end{figure}

\begin{figure}[h!]
\includegraphics[width=0.95\textwidth]{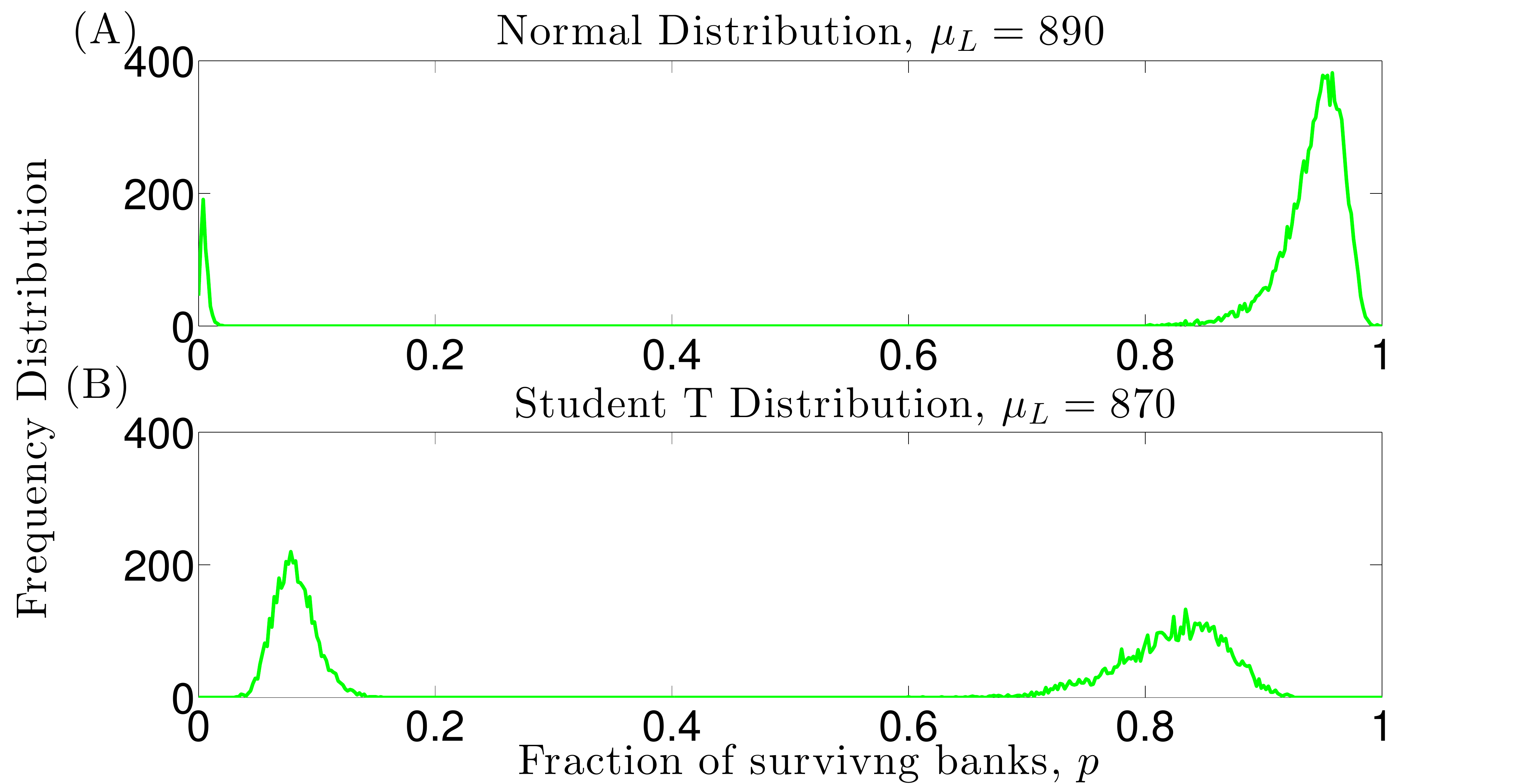}
\caption{This figure shows the frequency distribution of fractions of surviving banks $p$ for banks initialized with Normal (A) and Student's t distribution (B) for fixed values of $\mu_L, \mu_A, \sigma_L$ and $\sigma_A$. The fraction of interbank loans to total assets $\theta$ is set to 0.3 and the underlying structure of the exposure network is an Erd\H{o}s-R\'{e}ny network. To observe the behaviour in the proximity of the jump the values for $\mu_L$ where set to 890 for the Normal distribution and 870 for the Student's t distribution. 
To compute the frequency distribution, we repeated the simulation 10000 times. Two peaks occur because of perturbations of the balance sheet values due to the randomness. The two peaks are visible in both sub-plots at the end and beginning of the scale of $p$ indicating that most of the banks in the banking system either survive or are distressed. Intermediate fractions of surviving banks do not occur. }
\label{fig:survivaldistribution}
\end{figure}

The effects of different underlying distribution are illustrated in Figures~\ref{fig:DifferentDistributions} and~\ref{fig:survivaldistribution}. The underlying network structure of the exposure matrix is, in both figures, an Erd\H{o}s-R\'{e}ny network.

In Figure~\ref{fig:DifferentDistributions}, we report the average simulated fraction of surviving banks against $(\mu_L-\mu_a)/(\sigma_A^2+\sigma_L^2)^{1/2}$ and the fixed point solution of the Iteration Function~\ref{MeanFieldSurv} (black line) against $a-b$. For the simulated fraction, we varied $\mu_L$ and for the fixed point solution, we changed $a$ to satisfy $(\mu_L-\mu_a)/(\sigma_A^2+\sigma_L^2)^{1/2} \approx a-b$. For each $\mu_L$, the simulation was repeated 100 times. 
In the figure, symbols represent average fractions and vertical error bars are the standard deviations from the 100 simulations. To test the behaviour of the simulation for different fractions of average interbank loans, we changed $\theta$ from 0.0 (blue line), to 0.1 (red line) and 0.3 (green line). To compute the equivalent fixed point solution for each value of $\theta$, we changed the value for $b$ in the Iteration Function~\ref{MeanFieldSurv} accordingly, i.e. $b \approx \theta \mu_A/\sqrt{\sigma_A^2+\sigma_L^2}$. The critical value for $b$ for the normal distribution is $b_c=\sqrt{2\pi}$. For the Student's t distribution with 2 degrees of freedom, the critical value for $b$ is reached when $b_c \approx 2.82$. Hence, $\theta = 0.1$ leads to a value of interbank assets of bank $i$ below the critical value and, conversly, setting $\theta = 0.3$ creates a value of interbank assets above the critical value where a jump becomes visible. 

The difference between Figures~\ref{fig:iterationfunction2} and~\ref{fig:DifferentDistributions} is that to compute the fixed point solution in Figure~\ref{fig:iterationfunction2}, the total assets of the banks are varied as the mean of non-interbank assets is constant and a change in $b$ implies that either capital is changed to compensate a decrease or increase in total assets, or $\mu_L, \mu_g, \sigma_L$ and $\sigma_g$ change accordingly such that $a$ is constant. Whereas, in Figure~\ref{fig:DifferentDistributions}, the mean of the total assets of banks is constant and a change in $\theta$ does not effect the size of the balance sheet. Hence, capital stays constant for fixed values of $\mu_L, \sigma_L$ and $\sigma_g$.

The fractions of surviving banks computed in Figure~\ref{fig:DifferentDistributions} used Normal distributions ($A$) and Student's t distributions ($B$) to initialize total assets and total liabilities. Similarly, to compute the fixed point solutions, we used a standard normal CDF in $A$ and a standard Student's t CDF in $B$. 

We note that, for  $\theta = 0.3$ more banks default for the same values of  $\mu_A$, $\mu_L$, $\sigma_A$ and $\sigma_L$ than when $\theta = 0$. The reason is that there exists no counter party risk when $\theta = 0.0$. For both distributions a sudden decrease in the fraction of surviving banks is observed for $\theta = 0.3$. 
The jump starts earlier for the banking system with banks initialized with the Student's t distribution than for banks initialized with the Normal distribution. Also, the simulation results for a banking system initialized with Normal distributions are a closer fit to the fixed point solutions of the Iteration Function~\ref{MeanFieldSurv}, nonetheless the simulated results initialized with the Student's t distribution are also reasonable close to the fixed points.  
In the proximity of the jump, the standard deviation of the simulated fractions of surviving banks increases. This indicates that for the values of $\mu_A, \mu_L, \sigma_A$ and $\sigma_L$, at which the jump occurs, either most of the banks are operating or most of the banks are in distress with no intermediate state.

To investigate this behaviour for parameter values  close to the jump, we plotted the frequency distribution for fixed values of $\mu_A, \mu_L, \sigma_A$ and $\sigma_L$ in proximity of the jump in Figure~\ref{fig:survivaldistribution}. We used different values of $\mu_L$ for the simulations when initializing with Normal distributions ($\mu_L = 890$) and Student's t distribution ($\mu_L = 870$). This is because of the jump starting earlier for the Student's t distribution than for the Normal distribution. The value for $\theta$ is set to 0.3 again. To determine the frequency distribution, we repeated the default algorithm for the fixed values of $\mu_A, \mu_L, \sigma_A$ and $\sigma_L$ 10,000 times and sum the occurrence of the same equilibrium fraction of surviving banks. Sub-plot $A$ shows the results for simulations using the Normal distribution and sub-plot $B$ shows the results for simulations using the Student's t distribution.
For both distributions, two peaks occur. The peaks of the frequency distribution for a banking system initialized with the Normal distribution occur around $p$ close to zero and for $p$ between 0.9 and 1.0. The first peak for the fraction of surviving banks for a banking system with balance sheets initialized with the Student's t distribution happen between 0.01 and 0.15 and the second peak for values of $p$ between $0.65$ and $0.95$. Values of fractions of surviving banks between the two peaks do not occur. The lack of intermediate values is due to the stable and unstable fixed points. The unstable fixed point forms a barrier between the stable fixed points. However, slight perturbations of the values of banks assets and liabilities caused by the randomness of the simulation either tip the banking system into distress or survival. 

The number of banks defaulting before the sudden system failure happens when initialized with the Normal distribution is less than for a banking system initialized with the Student's t distribution. The Student's t distribution is a fat tail distribution implying that banks balance sheets differ more than when the balance sheet values are distributed with a Normal distribution. Thus, after the jump some banks have a greater chance of survival, as they have more capital, than other members of the banking system. However, because of the greater diversity, some banks also have less capital than other banks, causing the system failure to happen for a smaller mean value of liabilities in comparison to a more homogeneous banking system when initialized with the Normal distribution. Thus, the more diverse system is more prone to failure but chances of survival of some banks are larger than for a more homogeneous banking system. 

\subsection{Network Influence}\label{Sec:NetworkInfluence}

Interbank networks of various countries (Austria \cite{Boss2004}, Brazil \cite{Cont2010}, UK \cite{Langfield2012}, Italy \cite{Iori2008}, etc.) have been studied with the outcome that the networks do not  resemble Erd\H{o}s-R\'{e}ny networks. 
Instead, they consist of ``low clustering coefficients with short average path length " \cite{Boss2004} and the links in the interbank networks resembling the exposure from one bank to others are distributed with tails  exhibiting ``a linear decay in log-scale, suggesting a heavy Pareto tail" \cite{Cont2010} indicating a core-periphery structure with banks in the centre being highly connected and periphery banks being connected to the core banks \cite{Viegas2013}.

\begin{figure}[h!]
\includegraphics[width=0.95\textwidth]{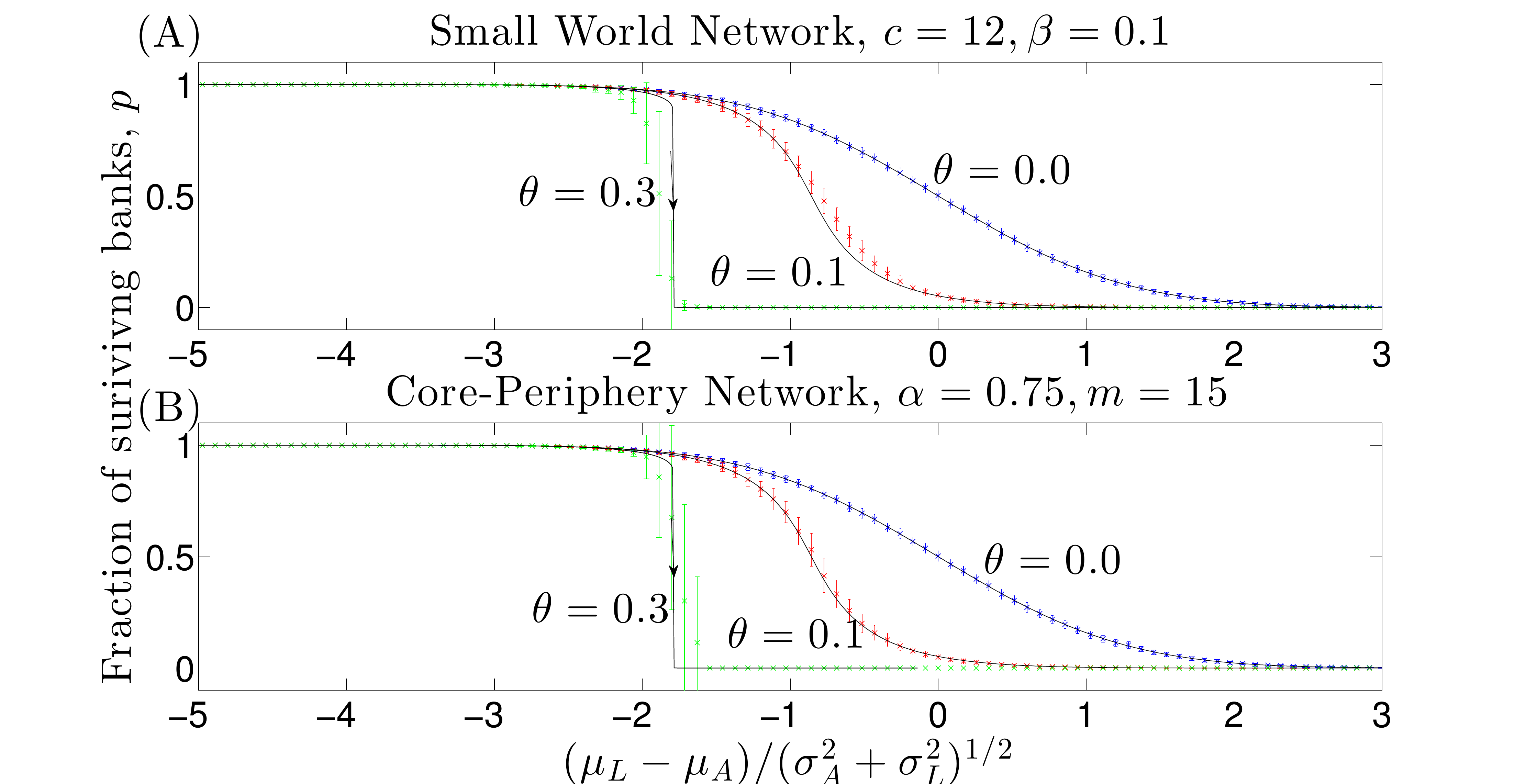}
\caption{The figure shows the average fraction of surviving banks $p$ computed using 100 simulations plotted against $(\mu_L-\mu_A)/(\sigma_A^2+\sigma_L^2)^{1/2}$. The balance sheet values are normally distributed. The underlying structure of the exposure networks are Small-World with neighbouring nodes $c=12$ and a re-wiring probability $\beta$ set to 0.1 (A) and core-periphery networks with a strongly connected cores created using Erd\H{o}s-R\'{e}ny  networks with connection probability $\alpha =0.75$ and 50 banks, and 450 periphery banks that are added one by one and joint to $50$ already existing banks using
the preferential attachment algorithm. As in Figure~\ref{fig:DifferentDistributions}, for a fraction of interbank assets to total assets, $p$ is plotted using green symbols, for $\theta = 0.1$ we used red symbols and for $0.0$ blue symbols were used. The error bar is the standard deviation of the results of 100 trials. The black line represents the fixed points of the Iteration Function~\ref{MeanFieldSurv} plotted against $a-b$ for changing $\theta$ as used in the simulation. The values of $p$ for the simulation and the Iteration Function~\ref{MeanFieldSurv} are for both network structures reasonable close and the steep decrease in the proximity of the jump are for both network structures observable. }
\label{fig:simulationNetworks2}
\end{figure}

In Figure~\ref{fig:simulationNetworks2}, we test the influence of other exposure network structures than the Erd\H{o}s-R\'{e}ny network. The distributions used to initialize the balance sheets for both sub-plots are Normal distributions. The structure of the outline of Figure~\ref{fig:simulationNetworks2} is similar to the one in Figure~\ref{fig:DifferentDistributions}. Again, we plotted the average fraction of 100 trials of surviving banks for a $\theta$ of 0.3 (green line), 0.1 (red line) and 0.0 (blue line) against $(\mu_L-\mu_a)/(\sigma_A^2+\sigma_L^2)^{1/2}$ varying $\mu_L$. The black lines are the solution of the fixed points of Eq.~\ref{MeanFieldSurv} for changing $b$ to match the equivalent value of $\theta$. Plot $A$ shows the results given that the underlying exposure network has a Small-World structure and in plot $B$, the underlying exposure network structure uses the preferential attachment algorithm to create a core-periphery structure. To tidily connected the core banks, we used Erd\H{o}s-R\'{e}ny core networks made out off 50 banks with a connection probability $\alpha$ of 0.75. The remaining 450 periphery banks are added one-by-one connecting to 15 banks using the preferential attachment algorithm.  

The simulation results using both network structures are reasonable close to the fixed point solutions of the Iteration Function~\ref{MeanFieldSurv} with a steep decline in surviving banks for $\theta =0.3$. 

The steep decline of $p$ when the Small-World network is used starts a bit earlier than the predicted jump in the mean-field. Before the rewiring process, the Small-World network is an ordered lattice. The Ising model on an ordered lattice can be approximated using the mean-field solution as long as the number of close neighbours is larger than 4. The re-wiring creates long-distance links between banks distributing the shock quicker through the network. 

Thus, it can be said that the network influence is marginal given that the number of lending banks is large enough. This can be explained using the results in Section~\ref{Sec:BF}. There, we showed that when $p_r = x_1$ (and assuming a small change from $p_r$ to $p_{r-1}$), the average number of banks failing as a result of one distressed bank is one again. Therefore, this implies that when capital is low the distress of one bank causes a chain of distress in connected banks resulting in distress throughout the entire system implying that the network structure is secondary. However, is has been reported that in the real world networks, periphery banks are of smaller size than core banks, which we did not account for and might lead to a different result.  

\subsection{Collaterals}\label{Sec:DerivativesandCollateral}

To incorporate collaterals of a lending agreement we add the following term to total assets, $A_i(r)$, of bank $i$ in round $r$:

\begin{equation}
\sum_{j=1}^M q\theta A_ig_{i,j}(1-S_j(r)),
\end{equation}  

where $q \in [0,\infty]$ accounts for the value difference of the loan from bank $i$ to bank $j$ and the collateral bank $j$ has to pay whenever it cannot pay its outstanding credit.

\begin{figure}[h!]
\includegraphics[width=0.95\textwidth]{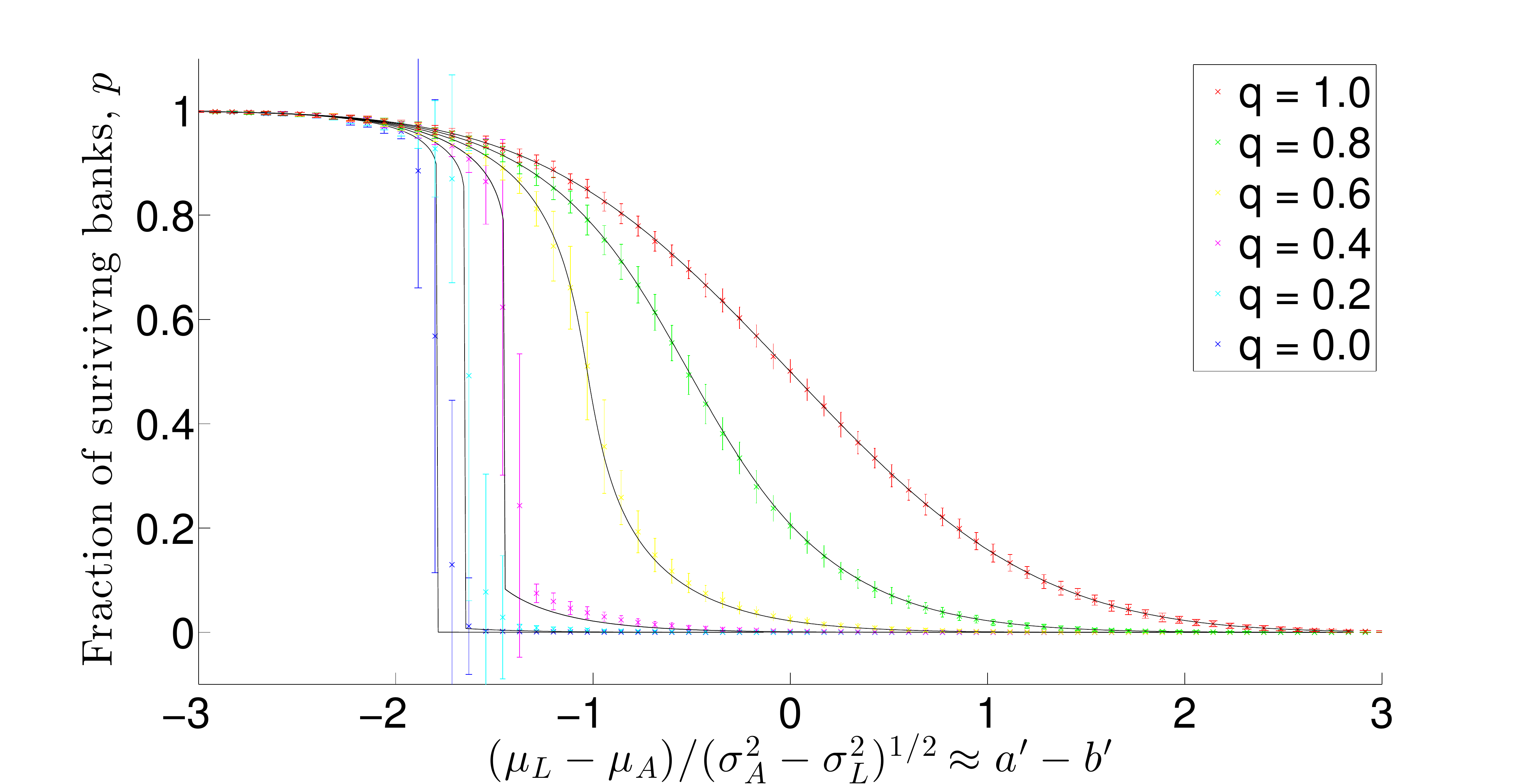}
\caption{The figure shows the average fraction of surviving banks $p$ computed using 100 simulations plotted against $(\mu_L-\mu_A)/(\sigma_A^2+\sigma_L^2)^{1/2}$. The balance sheet values are normally distributed. The underlying structure of the exposure network is an Erd\H{o}s-R\'{e}ny network. A collateral term was added when the total assets where computed during simulation modelling. The collateral on loans becomes active after the loaner defaulted. The black line represents the fixed points of the Iteration Function~\ref{MeanFieldSurv} using $a'$ and $b'$ as given in Eqns.~\ref{definitionOfaa} and~\ref{definitionOfbb} to compute the fixed point. The fixed points are plotted against $a'-b'$ ($\theta = 0.3$ was used in the simulation). The different coloured lines represent varying fractions $q$ such that the value of the collateral for any loan from bank $i$ to bank $j$ is $q\theta A_ig_{i,j}$. For increasing $q$ the interbank interaction is reduced such that for $q=1$ the interbank loans can be disregarded.}
\label{fig:collateral}
\end{figure}

Figure~\ref{fig:collateral} is a plot of the fraction of surviving banks $p$ using simulation testing including the collateral term and the fixed point solution of the Iteration Function~\ref{MeanFieldSurv} using $a'$ and $b'$ as given in Eqns.~\ref{definitionOfaa} and~\ref{definitionOfbb}. The average fraction of surviving banks was plotted for 100 trials along with the errorbars (coloured lines) for fixed $\theta = 0.3$. The black line are the fixed point solutions of the Iteration Function~\ref{MeanFieldSurv}. The different colours represent varying fractions of $q$. For increasing $q$ the interaction in form of the interbank loans between banks can be disregarded. However, for lower values of $q$ the jump can still be observed.    

\section{Analysis of real banking system data}\label{Sec:Data}

Banks report their balance sheet quantities yearly as part of their financial statement in their annual report. We used Bankscope \cite{Bankscope2011} to collect data for US and UK banks\footnote{The query settings were on "Status: Active Banks, Inactive Banks'', "Specialisation: Commercial banks, Savings banks, Cooperative banks, Real Estate \& Mortgage banks, Investment banks, Islamic banks, Other non banking credit institutions, Bank holdings \& Holding companies, Private banking $/$ Asset management companies'' and "Ultimate Owner: Def. of the UO: min. path of 50.01$\%$, known or unknown shareh., closest quoted company in the path leading to the Ultimate Owner (if any); GUO and DUO''}. The data includes consolidated values for some banks and unconsolidated values for others. Only using the values from consolidated balance sheets would have reduced the list of banks considerably mostly excluding foreign subsidiaries of foreign banks. We chose the years 2007 and 2012 as reference years, to determine the stability of the UK and US banking system during the recent financial crisis and a non-crisis time. The parameters $\mu_A$ and $\mu_E$ represent the "true'' of the average value of total assets and capital per bank.  

The two quantities that are decisive for the stability of the banking system in our model are the mean of the total assets $\mu_A = \mu_g + \theta \mu_A p_0$ (with $p_0 =1$) and the mean of loss absorbing capital $\mu_E = \mu_A-\mu_L$. We are using the "Tier 1 Capital'' and "Total Assets'' as reported in Bankscope. It should be noted that the UK and US use different accounting systems leading to different estimations for the value of the same asset and liabilities. Henceforth, the value of total assets, total liabilities and Tier 1 capital for UK and US banks reported in Bankscope cannot be compared country wise. However, it is possible to  discuss changes in financial stability of the banking systems in a country for different years.    To compute the mean values for $\mu_A$ and $\mu_E$ we only use banks with Tier 1 capital larger than zero this reduced the list of banks considerably (especially in 2007) as Bankscope does not report the Tier 1 capital value for all banks. The mean values as well as the number of banks used to compute the values can be found in Table~\ref{tab:data}. To compare the values for Tier 1 capital and total assets in the different years, we also included leverage, $\gamma$ in the table. It becomes clear that in      2007 the average leverage both in the US and UK was less than it was in 2012 and henceforth already implies a less stable system in 2007.

\begin{table}\caption{The table reports the mean value of total assets $\mu_A$ and Tier 1 capital of banks $\mu_E$ and the standard deviations for the years 2007 and 2012 for the UK and US banking system. The data is from Bankscope. We only considered banks that reported their Tier 1 capital. Thus, the table additionally states the number of banks. To compare the Tier 1 capital, we also stated the leverage ratio $\gamma$, i.e. Tier 1 capital to total assets.}
\begin{tabular}{ l l l l l  }
\hline\noalign{\smallskip}
   & \multicolumn{2}{c}{UK} &  \multicolumn{2}{c}{USA}\\
  \hline\noalign{\smallskip}
        & 2007 in GBP & 2012 in GBP & 2007 in USD & 2012 in USD \\ \hline\noalign{\smallskip}
$\mu_A$ & 2.0287e+11  & 1.8307e+11  & 1.8505e+10  & 2.0247e+10\\ 
STD     & 4.7503e+11  & 4.2912e+11  &1.3592e+11   & 1.5234e+11\\ 
$\mu_E$ & 6.3032e+09  & 8.1836e+09  & 1.0615e+09  & 1.5829e+09 \\ 
STD     & 1.3785e+10  & 2.0298e+10  & 6.6785e+09  & 1.1102e+10\\ 
Leverage, $\gamma$ & 0.0311 &0.0447 &     0.0574  &     0.0782 \\ 
No. banks & 26 & 38 & 666 & 779 \\ \hline\noalign{\smallskip}
\end{tabular}
\label{tab:data}
\end{table}
 
The parameter $\sigma$ is a free model parameter that indicates the uncertainty about the value of asset and liabilities.  More precisely $\sigma_g$ increases if the value for non-interbank assets is uncertain. Similarly, difficulties in obtaining funding from  banks or other funding sources are represented in an increased $\sigma_L$. In a way $\sigma$ is measures the severity of the shock and hence we tested for different values of $\sigma$. To calibrate $\sigma$, we use a variable $f \in [0,1]$ and say that $\sigma$ is a fraction of the mean value of the Tier 1 capital, $\mu_E$. Strictly speaking, $\sigma$ as discussed in the above analysis of the homogeneous banking model is the standard deviation of $\mu_L - \mu_g$, but as seen in Section~\ref{Sec:Sim} the difference between the standard deviation of $\mu_L - \mu_g$ and $-\mu_E = \mu_L - \mu_A$ is minimal. 

Another parameter that cannot be easily obtained from the annual account data is the average fraction of interbank assets, $\theta$. Banks report their lending to other banks under "Loans and advances to banks'' and "Deposits by banks'' in their annual reports. However, as it is pointed out in \cite{Langfield2012} loans and advances to banks are not the only exposure banks have to other banks. Such that to monitor the UK interbank market the Prudential Risk Authority (PRA) collects data about other financial instruments that form part of the interbank market. In particular \cite{Langfield2012} list: "prime lending (...); holdings of capital and fixed-income securities issued by banks; credit default swaps bought and sold; securities lending and borrowing (...); repo and reverse repo (...); derivatives exposure (...); settlement and clearing lines; asset-backed securities; covered bonds; and short-term lending with respect to other banks and broker dealers''.  The balance sheet data reported in the annual reports does not differentiate between the interbank market and products obtained from other financial institutions. Still using only the values for "Loans and advances to banks'' or "Deposits by banks'' to calibrate $\theta$ would underestimate the average fraction of interbank lending. Henceforth, we again use multiple values of $\theta$ to test the stability of the system.   

\begin{figure}[h!]
\includegraphics[width=0.95\textwidth]{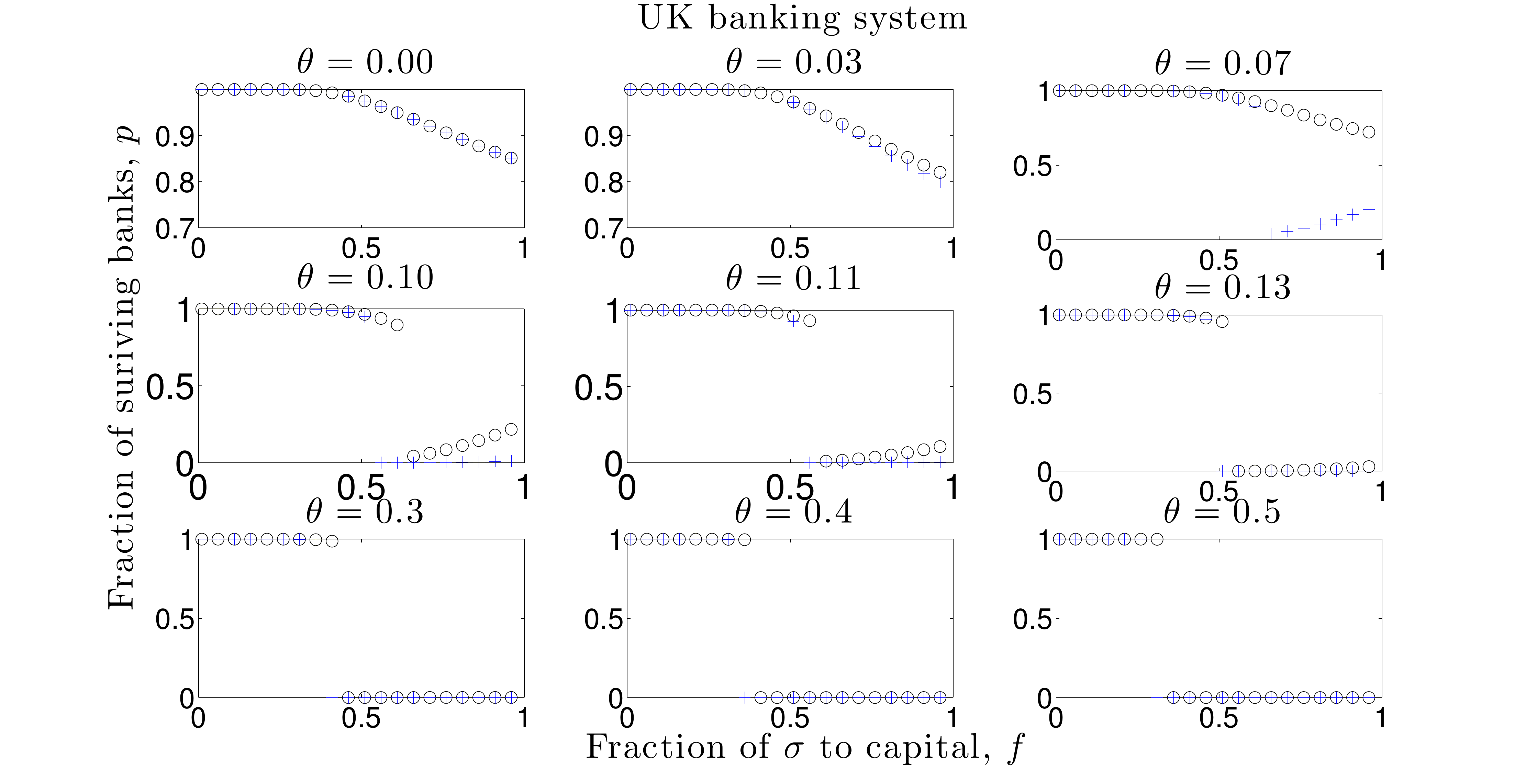}
\caption{The sub plots show the fraction of surviving banks for the years 2007 (blue crosses) and 2012 (black circles) against the fraction of $\sigma$ to mean value of capital, f, for various values of the fraction of interbank assets to total assets, $\theta$. To calibrate the model, the mean of total assets, $\mu_A$, and the mean of Tier 1 capital, $\mu_E$, was used from banks from the UK banking system. For $\theta = 0$, banks are not interconnected. In that case, for both years no  systemic distress event happens. In order for a system failure to happen, $\theta$ needs to be non-zero. The sudden system failure happens for the banking system calibrated with the 2007 UK data for $\theta = 0.07$ at which the banking system calibrated with 2012 UK data is still in a stable state. For $\theta \geq 0.10$, the banking system calibrated with 2012 UK data also becomes unstable for a large enough $f$. However, $f$ at which the systemic distress happens for the 2007 UK data is smaller then the value for $f$ at which the systemic failure happens when the banking system is calibrated with the 2012 UK data implying that the 2007 system is more prone to failure then the 2012 banking system.}
\label{fig:UK20072013}
\end{figure}

\begin{figure}[h!]
\includegraphics[width=0.95\textwidth]{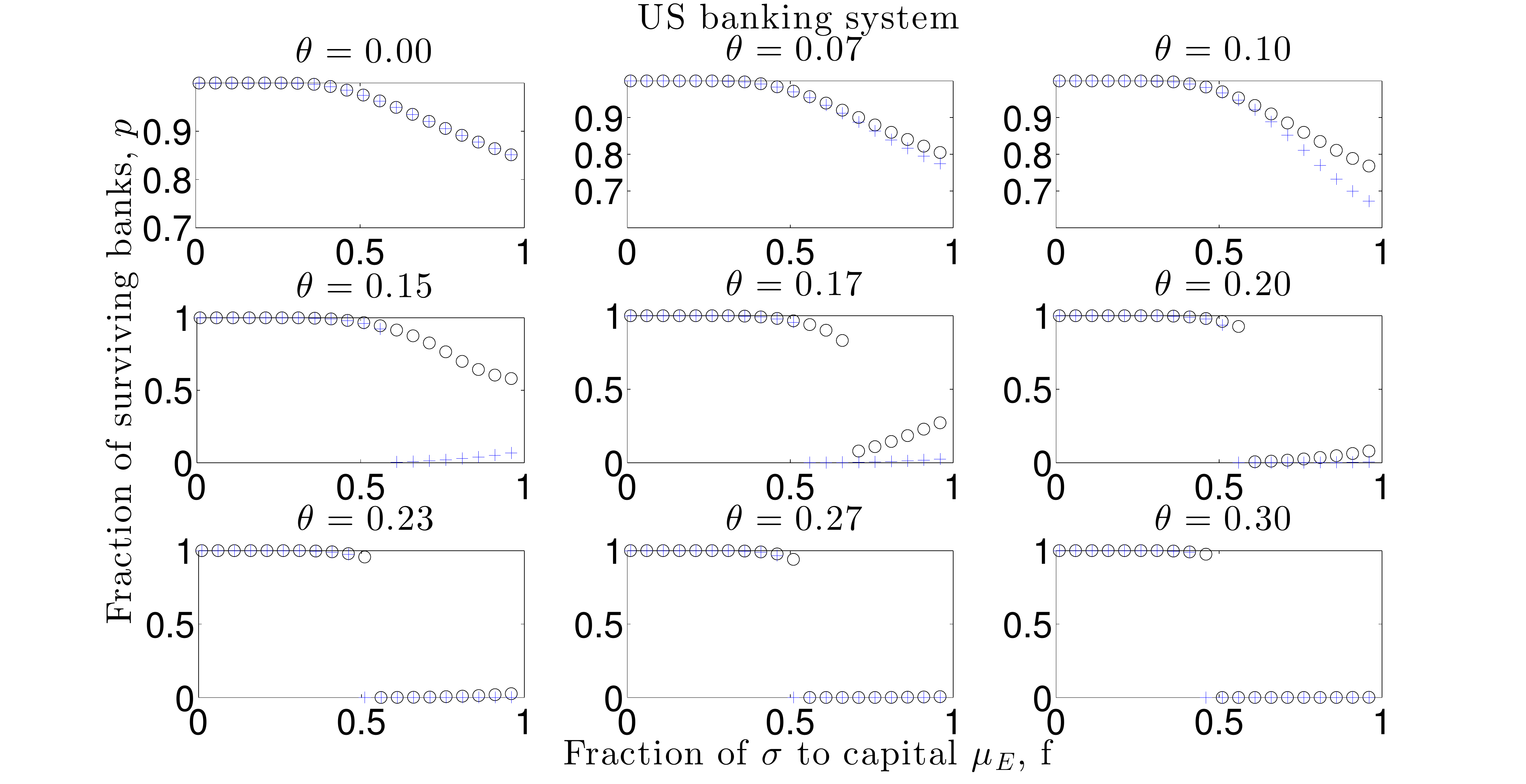}
\caption{The figure is similar to Figure~\ref{fig:US20072013} except that US balance sheet data for the years 2007 and 2009 was used to calibrate the model. The sub plots show the fraction of surviving banks for the years 2007 (cross) and 2012 (circle) against the fraction of $\sigma$ to mean value of capital, f, for various values of the fraction of interbank assets to total assets, $\theta$. To calibrate the model, the mean of total assets, $\mu_A$, and the mean of Tier 1 capital, $\mu_E$, was used from banks from the US banking system. For $\theta = 0$, banks are not interconnected. In that case, for both years no  systemic distress event happens. In fact even for an increased $\theta$ of 0.10 the system is stable with only some losses for large $f$ but no system-wide failure. The sudden system failure happens for the banking system calibrated with the 2007 US data for $\theta = 0.15$. However, we note that for the same value of $\theta$, the banking system calibrated with 2012 US data is still in a stable state. For $\theta \geq 0.17$, the banking system calibrated with 2012 UK data also becomes unstable for a large enough $f$. For both years, $\sigma$ needs to be at least half of the size of banks capital in order for the system wide failure to happen.}
\label{fig:US20072013}
\end{figure}

Figures~\ref{fig:UK20072013} and~\ref{fig:US20072013} show various plots the fraction of surviving banks, $p$, plotted against the fraction of $\sigma$ to the mean Tier 1 capital $\mu_E$, $f$ for the UK and US system, respectively. The fraction of surviving banks is calculated using the fixed points of Eq.~\ref{MeanFieldSurv} using a standard normal CDF as before. The value of the fraction of interbank lending to total assets, $\theta$ is fixed and given above each sub plot. The blue crosses indicate the fraction of surviving banks for a banking system calibrated with the 2007 data and the black circles symbolizes the fraction of surviving banks for a banking system calibrated with the 2012 data. 

For $\theta$ set to zero the fraction of surviving banks in the UK banking system is almost identical (Figure~\ref{fig:UK20072013}). The number of surviving banks declines for a larger $f$. However, even for $f$ tending to one more than 85\% of banks are operating in both 2007 and 2012. Note that $\theta$ equal to zero corresponds to no interbank lending. The number of distressed banks is only due to the uncertainty of the value of liabilities and non-interbank assets caused by a large $\sigma$. For the range of $\sigma$ from zero to the size of $\mu_E$, no systemic event, i.e. the entire failure of the banking system, becomes possible in both years given that there is only a shock to the value of non-interbank assets or liabilities.   

For the next graphs in Figure~\ref{fig:UK20072013} in the first row $\theta$ is increased to 0.03 and 0.07. It becomes clear that the fraction of surviving banks deviates for 2007 and 2012 with $p$ for 2007 being considerable less than $p$ for 2012 implying that the banking system 2007 was much more prone to failure. For $\theta = 0.07$ and the banking system calibrated with the 2007 data set, a jump becomes visible for $p$ for $f$ around 0.5. The banking system calibrated with the 2012 data set remains stable for $\theta$ set to either 0.03 or 0.07. This changes when $\theta$ is further increased. In the second row of Figure~\ref{fig:UK20072013}, $\theta$ is set to 0.10, 0.11 and 0.13. The sudden jump for banks calibrated with the 2007 data set happens for $f$ around 0.56 to 0.51 and increases even further in the third row when $\theta$ takes the values 0.3, 0.4 and 0.5 with a value of $f$ around 0.46 - 0.31 being sufficient to ensure an unstable banking system. For the banking system calibrated with the 2012 data set a jump also occurs for values of $\theta$ above and including 0.1. For $\theta$ equal to 0.10 the jumps happens for $f$ around 0.66. As for the 2007 data set, the jump moves to a lower value of $f$ for a larger $\theta$ with $\theta$ set to 0.5, $f$ being around 0.36 for the jump to happen. 

\begin{figure}[h!]
\includegraphics[width=0.95\textwidth]{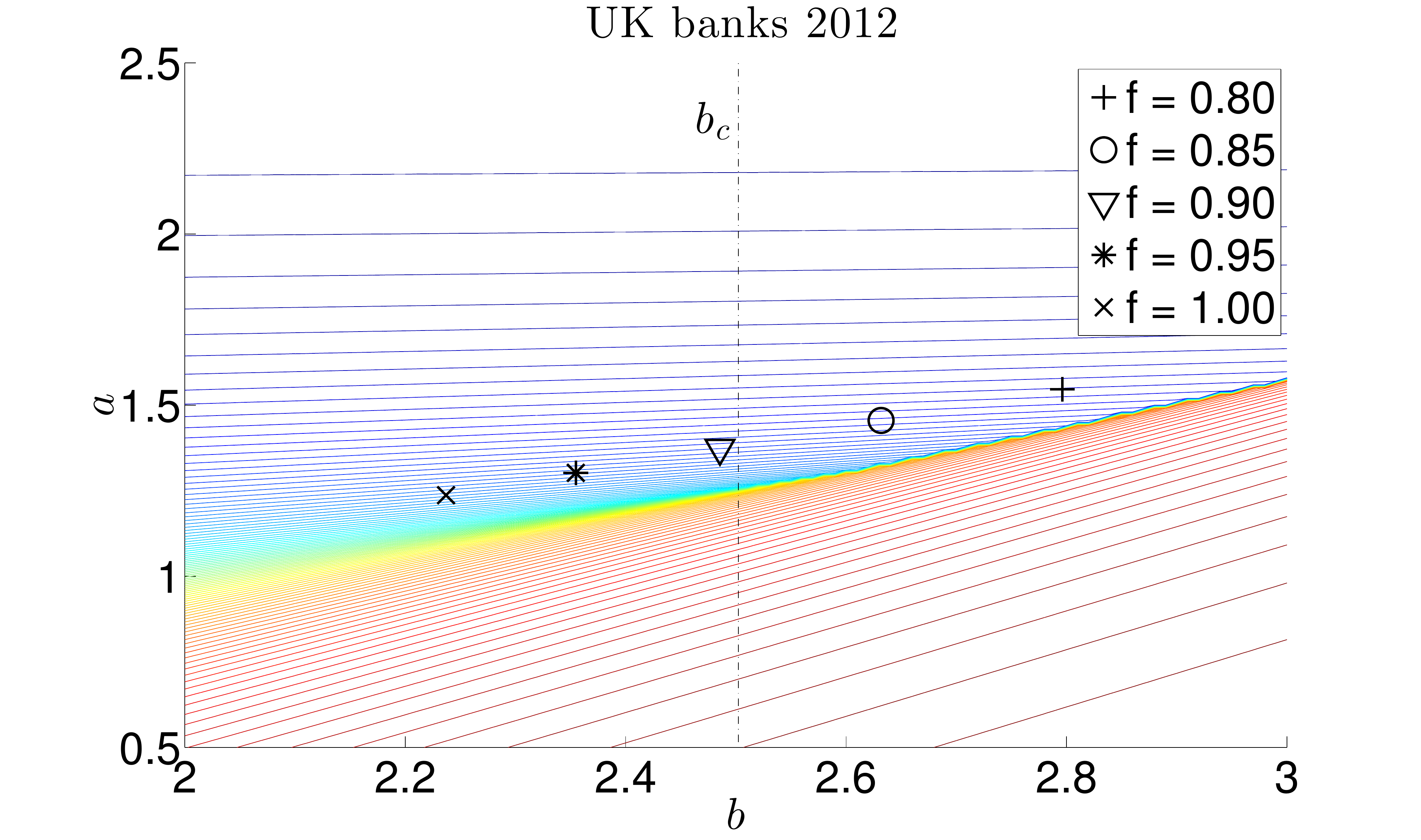}
\caption{The figure is similar to Figure~\ref{fig:phasespace3} showing the fraction of surviving banks for different values of $a$ and $b$. Additionally to the fraction of surviving banks for particular values of $a$ and $b$, we also plotted the particular values of the fraction of surviving banks calibrated with the 2012 UK data for $\theta$ fixed at 0.10 for varying $f$ as indicated. It becomes clear that for increasing $f$, $b$ decreases such that for $f = 0.90$, $b$ becomes less than $b_c$. Simultaneously, $p$ increases explaining the increase in $p$ observed in Figure~\ref{fig:UK20072013} for $\theta = 0.10$ and $\theta = 0.11$ for increasing $f$ for the 2012 UK data.}
\label{fig:regionOfStabilityUK2012theta010}
\end{figure}

For $\theta$ equal to 0.10 or 0.11 a jump occurs as well for the banking system calibrated with the 2012 UK data set. However, after the jump, $p$ increases for increasing $f$. This can be explained using Figure~\ref{fig:regionOfStabilityUK2012theta010}. Figure~\ref{fig:regionOfStabilityUK2012theta010} is the same plot of the contour lines of surviving banks as plotted in Figure~\ref{fig:phasespace3}. The black symbols indicate the position of $p$  for fixed $\theta$ equal to 0.10 and varying $f$ as indicated in the accompanying legend. It becomes obvious that for increasing $f$, $b$ decreases such that for $f = 0.90$ a jump does not become possible any more and the system is in the reversible region. At the same time, the value of $p$ increases for decreasing $b$. Hence, we can observe an increase in $p$ even so $f$ and henceforth the uncertainty $\sigma$ increases.   

Figure~\ref{fig:US20072013} is similar to Figure~\ref{fig:UK20072013} except that we used US banks to calibrate the model with the blue crossed line representing the fraction of surviving banks in 2007 and the black circled line being the fraction of surviving banks in 2012. In Figure~\ref{fig:US20072013} the difference in the stability of the US banking system in 2007 and 2012 is less visible suggesting that a shock of similar size as happened in 2007 would also cause severe damage in 2012. 

Figures~\ref{fig:UK20072013} and~\ref{fig:US20072013} show that exposure to other banks played an important role in the recent financial crisis. As we mentioned before we cannot be certain about the actual average fraction of interbank loan nor the size of $\sigma$ at that time.  However, an exposure of 30\% of total assets to other banks seems like a valid estimate. A $\sigma$ of 25\% or 50 \% of the bank's capital only happens during a period of large uncertainty - which one can argue happened during the 2008 meltdown of the financial sector. In particular, the Financial Services Authority (FSA) stated in their report on "The failure of the Royal Bank of Scotland'' \cite{fsa2011} that beside mismanagement a mismatch in short-term funding and devaluation of long-term assets played part of the failure and eventual bail-out of the Royal Bank of Scotland by the UK government. In the 2012 data set, for 30\% interbank assets to total assets, $\sigma$ needs to be much larger for the jump to occur implying a more stable system. This is due to more capital in the banking system. Needless to say that using the balance sheet test to determine insolvency, a bank failure is always an option as capital is limited. The likelihood of such a large shock to happen is not part of this paper but certainly it can be considered a rare event. Nonetheless, the maximal economical feasible leverage ratio should be used as a minimum to prevent entire system failure and taxpayer intervention. 
   
\section{Conclusions}
\label{Sec:Con}

We studied a stylized banking model based on balance sheet quantities to understand the influence of counter party failure on the stability of the entire banking system. In our stylized banking model, the number of bank failures can be reduced by increasing the amount of lending in the interbank system which can compensate for fluctuations in the assets and liabilities. However, above a certain critical fraction of the average amount borrowed with respect to the average combined fluctuations in liabilities and non-interbank assets a single bank failure can trigger catastrophic events that can bring down the entire system. In this regime, the system is irreversible and the normal operating state can only be recovered at a cost of introducing capital externally. In addition to estimating the cost of rescuing a failed banking system, we stated a minimum leverage requirement that ensures a stable system. We have archived this by solving a fixed point equation that reveals at the transition two stable fixed points separated by a barrier in the form of an unstable fixed point.

We archived this by constructing a Merton model of default where banks interact with one another via interbank lending assigning banks a state whether they are normally operating or are in distress. This allowed us to use a stability analysis of the fixed points to investigate the stability of the banking system. The model uses balance sheet quantities to determine the counter party risk of banks. The initial round of distressed is caused by changes in the distributions of non-interbank assets and liabilities. We showed that depending on the balance sheet parameters, non, partial, and entire failure due to counter party risk of the banking system becomes possible. 

We have verified numerically that this behavior is robust for different kinds of distribution of assets and liabilities fluctuations and for different types of interbank networks. We used simulations distributing assets and liabilities randomly among banks varying the average capital and thereby creating the initial round of default. The initial round of default created subsequent defaults caused by reduction of the total assets due to distressed counter parties.     We showed that the predicted jump indeed occurs for different distributions and various network structures. 

Finally, we used balance sheet data of UK and US banks from the years 2007 and 2012 to demonstrate the stability of the banking systems in the individual years. We showed that interbank lending made both the US and UK systems more prone to failure in 2007 such that small fluctuations in assets and liabilities could have caused catastrophic events. In 2012 for the same fluctuations both banking systems are still stable with much larger fluctuations needed to create a system-wide bank failure. 

We would like to stress that the numbers evaluated with the model should be taken with caution as the model is a simplification of real world events - as any model always will be. Also by no means are we claiming that the underlying distribution used to evaluate the stability of the model is a Standard Normal distribution or Student's t distribution. However, it explains the propagation of distress in a connected banking system and explains the mechanism and conditions under which a system failure occurs. The simple model of banking failure demonstrates the risk that counter party failure imposes in a highly connected banking system and is an explanation as to why the recent financial crisis had such a big impact even if it started with a local shock in the US mortgage market. 

An advantage of our model is, that interactions between variables can be included. For example, we tested the effects of collateralized lending but the impact of credit derivatives insuring against counter party risk could also be included. Except of changing the underlying exposure matrix, we did not explore the effects of heterogeneity of banks on the stability of the system. The analysis of the network structure suggests that for banks of similar size and exposed to similar market risk, the interbank network is not important. However, real world interbank networks are structures such that periphery banks are mostly small regional banks with core banks being internationally operating banks. Clearly, the default of a regional bank will not have the same impact as the default of an internationally operating bank. However, this effect is not covered in our homogeneous model. Similarly, discussions about ring-fencing the banking system or using the Volcker rule to separate investment and retail banking suggest that a shock to specific asset classes might not be as severe to specific kind of banks as to others and henceforth reduce the overall shock to the system. Changing the above model to a heterogeneous system might give answers to some of these questions.   

\begin{acknowledgements}
Support of the Economic and Social Research Council (ESRC) in funding the Systemic Risk Centre is acknowledged (ES/K002309/1). 
\end{acknowledgements}

\bibliographystyle{spmpsci}      

\nocite{*}
\bibliography{biblio}

%
%

\end{document}